\documentclass[prd,twocolumn,superscriptaddress,longbibliography,nofootinbib,notitlepage]{revtex4-1}

\usepackage{amsmath,amssymb,amsfonts,mathrsfs}
\usepackage{graphicx}
\usepackage{xcolor}
\usepackage{braket}
\usepackage{booktabs}
\usepackage{bm}
\usepackage{hyperref}

\newcommand{\vp}{\mathrm{vp}}
\newcommand{\phisq}{\langle\phi^2\rangle}
\newcommand{\mr}{\mathcal{R}}
\newcommand{\g}{\mathscr{G}}
\newcommand{\half}{\tfrac12}

\begin{document}

\title{Vacuum Polarization in Quantum-Corrected and Effective Black Hole Geometries: a High Performance Approach}

\author{Antonino Flachi}
\email{flachi@keio.jp}
\affiliation{Department of Physics, \& Research and Education Center for Natural Sciences, Keio University, 4-1-1 Hiyoshi, Yokohama, Kanagawa 223-8521, Japan}

\date{\today}

\begin{abstract}
We compute the renormalized scalar vacuum polarization $\phisq$ of a massive,
non-minimally coupled quantum field in the Hartle--Hawking state exterior to a
family of static, spherically symmetric \emph{quantum-corrected},
\emph{effective} and \emph{regular} black holes: the Kazakov--Solodukhin (KS)
quantum-deformed black hole, the two effective loop-quantum-gravity geometries of
Zhang--Lewandowski--Ma--Yang (ZLMY-1 and ZLMY-2), the renormalization-group
(RG) improved Schwarzschild black hole of Bonanno--Reuter (BR), and the
Bardeen regular black hole.
We adapt the extended point-splitting formalism, implemented as
a high-performance code, and present the exterior profile
$\vp\equiv16\pi^2M^2\phisq$ over a grid of the quantum-deformation parameter, the
field mass $\mu=mM$ and the curvature coupling $\xi$. KS \emph{suppresses} the near-horizon polarization, while the four
geometries with negative horizon Ricci scalar \emph{enhance} it. Measured
against the horizon curvature $M^2\mr_{\rm H}$, a measure of the
deformation with the same meaning in every theory, the $\xi$-dependent part
of the response, set by the DeWitt--Schwinger term linear in
$(\xi-1/6)\mr$, is common to all five geometries to within $\pm20\%$ at
light field mass, and at
equal curvature the total responses of ZLMY-1, BR and Bardeen agree to within a
factor $1.6$; a further, $\xi$-independent part set by the horizon data is
geometry specific. We verify the curvature mechanism both in magnitude and in
the $\xi$-dependence. In the near-extremal regime of the two geometries
with inner horizons we find a sign change of the horizon polarization at light
field mass, appearing only once the horizon Ricci scalar has turned
positive, removed by conformal coupling, and not found for
Reissner--Nordstr\"om ($\mr=0$) at matched temperature or matched horizon
separation in the range probed, which identifies it as the
DeWitt--Schwinger response to a positive horizon curvature. Every result reduces to the Schwarzschild value in
the classical limit, validating the calculation. These constitute, to our
knowledge, the first determinations of $\phisq$ for the KS, ZLMY, BR
and Bardeen black hole geometries, providing a quantitative measure of quantum-field activity on these effective backgrounds.
\end{abstract}

\maketitle

\section{Introduction}
\label{sec:intro}
The behavior of quantum fields in the exterior of quantum-corrected black holes
is a sensitive probe of the underlying short-distance physics: corrections of
quantum-gravitational origin can leave imprints at macroscopic scales, which,
within the effective-geometry framework adopted here (Sec.~\ref{sec:geometries}),
are transmitted to quantum fields through the metric.
The simplest local, gauge- and state-dependent diagnostic of this activity is the scalar vacuum polarization
$\phisq$, the renormalized coincidence limit of the field two-point function. Its
computation in a black hole background is a long-standing technical
problem. The extended point-splitting formalism~\cite{Candelas:1980,DeWitt:1965d,Christensen}, refined over
several decades~\cite{Anderson:1990jh,Winstanley:2008sr,Flachi:2008sr,Quinta:2016,Flachi:2015,Morley:2021},
provides a controlled route, and methodologically distinct schemes have also been developed, e.g.,~\cite{Breen:2016,Breen,Thompson}.

In an earlier paper~\cite{Flachi:2026} we have considered the problem of calculating the vacuum polarization for the loop-quantum-gravity black hole solution of Refs.~\cite{AO1,AO2,AO3}, as the simplest possible example to implement a high performance numerical scheme of the extended point-splitting approach. Here, we extend the computation to a broader family of \emph{uncomputed} geometries, each representing a different mechanism by which quantum or effective dynamics deform the Schwarzschild solution:
\begin{itemize}
\item the \textbf{Kazakov--Solodukhin (KS)} black
  hole~\cite{KazakovSolodukhin}, a two-dimensional-dilaton-inspired quantum
  deformation that removes the central singularity, controlled by a length
  $a$;
\item the two \textbf{Zhang--Lewandowski--Ma--Yang (ZLMY)} effective
  loop-quantum-gravity black holes~\cite{ZLMY}, controlled by a parameter
  $\zeta$; ZLMY-1 deforms both metric potentials while ZLMY-2 deforms only the
  radial one, so that $g_{tt}g_{rr}\neq-1$, i.e., lapse and radial functions are deformed independently of
  each other;
\item the \textbf{Bonanno--Reuter (BR)} renormalization-group-improved
  Schwarzschild black hole~\cite{BonannoReuter}, in which the Newton constant
  runs with distance, producing a Reissner--Nordstr\"om-like two-horizon
  structure governed by a single dimensionless parameter $\Omega$;
\item the \textbf{Bardeen} regular black hole~\cite{Bardeen1968,AyonBeato2000},
  a magnetically charged nonlinear-electrodynamics geometry with a regular
  (de Sitter) core, controlled by the dimensionless charge $q\equiv g_{\rm m}/M$; we
  treat the \emph{black-hole} (two-horizon, Hartle--Hawking) regime
  $q<q_c=4/(3\sqrt3)\simeq0.770$, distinct from the horizonless case of
  Ref.~\cite{BoassoMazzitelli}.
\end{itemize}
To the best of our knowledge $\phisq$ has not been computed for any of the above backgrounds: the existing literature treats their quasinormal modes, shadows, thermodynamics and Hawking emission, but not renormalized local expectation values of quantum fields.
For the Bardeen geometry $\phisq$ was very recently obtained in the
\emph{horizonless} case~\cite{BoassoMazzitelli}; here we treat instead the
black-hole regime, with an outer Killing horizon and the Hartle--Hawking thermal
state.

We follow closely the methodology of Refs.~\cite{Flachi:2008sr,Quinta:2016,Flachi:2026}, to which we refer
for full details, and only summarize it in Sec.~\ref{sec:formalism}. The central
new technical element is a partially \emph{metric-agnostic} implementation: the lapse and
radial functions $f$ and $g$, together with the analytic derivatives entering the
counterterms, are generated symbolically from a configuration file, so that a
single validated code base treats every geometry on the same footing and produces
a uniform data set. Section~\ref{sec:geometries} defines the geometries and
records the quantities (horizon location, surface gravity, Ricci scalar) that
enter and validate the calculation; Sec.~\ref{sec:formalism} summarizes the
formalism; Sec.~\ref{sec:implementation} describes
the implementation and its deployment on high-performance-computing
resources; Sec.~\ref{sec:results} presents the results;
Sec.~\ref{sec:nearext} pushes the two geometries with an extremal limit into
the near-extremal regime; we conclude in Sec.~\ref{sec:conclusions}.

Throughout we set $G_0=c=\hbar=1$, work in units of the Schwarzschild radius
$r_s=2M$, write $x\equiv r/r_s$, and report the dimensionless combination
$\vp\equiv16\pi^2M^2\phisq$ as a function of $x$, quoting the field mass through
$\mu\equiv mM=m/2$. The massless Schwarzschild horizon value is $\vp=1/12$.

\section{Quantum-corrected geometries}
\label{sec:geometries}
All five backgrounds are static and spherically symmetric,
\begin{equation}
ds^2=-f(r)\,dt^2+\frac{dr^2}{\,\tilde g(r)\,}+r^2 d\Omega_2^2 ,
\label{eq:lineelement}
\end{equation}
and we use the ``engine'' variables $f(x)$ and $g(x)\equiv x^2/\tilde g(x)$, in which
the Schwarzschild solution is $f=1-1/x$, $g=x^2/f$. We list each lapse, its
classical ($\to$ Schwarzschild) limit, the outer horizon $x_+$, the surface
gravity $\kappa$ and the Ricci scalar $\mr$. The latter is non-vanishing for all
five (none is a vacuum solution) and drives the DeWitt--Schwinger curvature
response discussed in Sec.~\ref{sec:results}.

\paragraph*{Kazakov--Solodukhin (KS).}
With $a$ a quantum-deformation length and $a_{\rm KS}\equiv a/r_s$,
\begin{equation}
f_{\rm KS}(x)=\frac{\sqrt{x^2-a_{\rm KS}^2}-1}{x},\qquad
g_{\rm KS}(x)=\frac{x^2}{f_{\rm KS}(x)} .
\end{equation}
The horizon sits at $x_+=\sqrt{1+a_{\rm KS}^2}$ and the surface gravity
$\kappa=1/(2r_s)$ is \emph{independent} of $a_{\rm KS}$: KS shares the
Schwarzschild Hawking temperature at any deformation and has no extremal limit.
The Ricci scalar is positive in the exterior and is $O(a_{\rm KS}^4)$: the
$O(a_{\rm KS}^2)$ truncation of the lapse,
$f=1-1/x-a_{\rm KS}^2/(2x^2)+O(a_{\rm KS}^4)$, is precisely of
Reissner--Nordstr\"om form and therefore has $\mr=0$ exactly, so the
deformation first enters $\mr$ at $O(a_{\rm KS}^4)$, where the expansion gives, to leading order,
$\mr=\tfrac34\,a_{\rm KS}^4/x^6$; $a_{\rm KS}\to0$ recovers Schwarzschild.

\paragraph*{Zhang--Lewandowski--Ma--Yang (ZLMY).}
Defining $F(x)=(1-1/x)\,[\,1+(\zeta^2/x^2)(1-1/x)\,]$, the two effective
geometries are
\begin{align}
\text{ZLMY-1:}\quad & f=F,\qquad g=x^2/F, \\
\text{ZLMY-2:}\quad & f=1-\tfrac1x,\qquad g=\frac{x^2}{F}\;\big(\neq x^2/f\big).
\end{align}
ZLMY-1 deforms both $g_{tt}$ and $g_{rr}$ (so $g_{tt}g_{rr}=-1$), whereas ZLMY-2
deforms only $g_{rr}$, making it the first case in which the two metric
functions are deformed independently of one another (the
background of Ref.~\cite{Flachi:2026} has $g_{tt}g_{rr}\neq-1$ only
through the fixed Lifshitz factor $r^{2\epsilon}$). For both, the bracket equals unity at $x=1$, so the horizon is
$x_+=1=r_s$ and $\kappa=1/(2r_s)$ \emph{independently} of $\zeta$ -- again no
extremal regime. For ZLMY-2, where $f\neq\tilde g$, the regularity of the
Euclidean section at the horizon -- on which the Hartle--Hawking
construction rests -- is not automatic and deserves an explicit
statement: smoothness requires $f$ and $\tilde g$ to vanish linearly at
the \emph{same} point, whereupon the near-horizon Euclidean geometry is
$ds_E^2\simeq\kappa^2\rho^2\,d\tau^2+d\rho^2+r_+^2\,d\Omega_2^2$, with
$\rho$ the proper radial distance and
$\kappa=\tfrac12\sqrt{f'(x_+)\,\tilde g'(x_+)}$, and the period
$\beta=2\pi/\kappa$ removes the conical singularity. ZLMY-2 satisfies
this: $F=f\,[1+(\zeta^2/x^2)f]$ has a simple zero at $x_+=1$ with
$\tilde g'(x_+)=f'(x_+)$ exactly (the bracket equals unity there), so the
horizon is a regular point of the Euclidean geometry and the thermal
state is well defined at $\beta=2\pi/\kappa=4\pi r_s$. Both are asymptotically flat with a leading $\zeta^2/x^2$
correction. The Ricci scalars are, in units $r_s=1$,
$\mr=2\zeta^2(2x-3)/x^6$ (ZLMY-1) and $\mr=\zeta^2(4x^2-14x+9)/(2x^6)$
(ZLMY-2): both are \emph{negative} at the horizon ($-2\zeta^2$ and
$-\zeta^2/2$, respectively) and turn positive farther out (beyond $x=3/2$
and $x\simeq2.65$); $\zeta\to0$ recovers Schwarzschild.

\paragraph*{Bonanno--Reuter (BR).}
RG improvement promotes $G_0\to G(r)$, giving the
lapse~\cite{BonannoReuter}
$f=1-2G_0M r^2/[\,r^3+\tilde\omega G_0(r+\gamma G_0 M)\,]$ with
$\tilde\omega=118/(15\pi)$ fixed. In our units, with
$\Omega\equiv\tilde\omega/(G_0M^2)$ a single dimensionless (mass-controlled)
parameter and $\gamma=9/2$ the physical value,
\begin{equation}
f_{\rm BR}(x)=1-\frac{x^2}{\,x^3+\tfrac{\Omega}{4}x+\tfrac{\gamma\Omega}{8}\,},
\qquad g_{\rm BR}=\frac{x^2}{f_{\rm BR}} .
\end{equation}
Unlike the previous cases, BR is Reissner--Nordstr\"om-like: $f$ has an outer and
an inner horizon that merge at an extremal value
$\Omega_{\rm cr}(\gamma=9/2)\simeq0.204$, and the surface gravity
\emph{decreases} with $\Omega$. In the main survey of
Sec.~\ref{sec:results} we stay safely sub-extremal,
$\Omega\le0.12\approx0.6\,\Omega_{\rm cr}$, avoiding the low-temperature
regime; the near-extremal limit $\Omega\to\Omega_{\rm cr}$ is explored
separately in Sec.~\ref{sec:nearext}.
The asymptotics are $f\to1-1/x+\Omega/(4x^3)$ (a $1/x^3$ correction, gentler than
RN), and over the survey range $\Omega\le0.12$ the Ricci scalar is
\emph{negative} throughout the exterior, $\mr\sim-\Omega/(2x^5)$; close
to extremality the horizon value turns positive
($\mr_{\rm H}r_s^2=-0.11$ at $\Omega/\Omega_{\rm cr}=0.99$, $+0.20$ at
$0.999$; Sec.~\ref{sec:nearext}); $\Omega\to0$ recovers Schwarzschild.

\paragraph*{Bardeen.}
The Bardeen regular black hole is the static, spherically symmetric solution of
gravity coupled to a magnetic-monopole nonlinear electrodynamics, with a single
metric function ($g_{tt}g_{rr}=-1$),
\begin{align}
f_{\rm Bardeen}(x)&=1-{x^2}/{\big(x^2+(q/2)^2\big)^{3/2}}, \\
g_{\rm Bardeen}&={x^2/ f_{\rm Bardeen}},
\end{align}
with $q\equiv g_{\rm m}/M$ the dimensionless charge ($g_{\rm m}$ the monopole
length). Like BR it
is Reissner--Nordstr\"om-like, with an inner Cauchy and an outer Killing horizon
that merge at the extremal value $q_c=4/(3\sqrt3)\simeq0.770$; in the main
survey of Sec.~\ref{sec:results} we stay sub-extremal,
$q\le0.6\approx0.78\,q_c$ (the near-extremal regime $q\to q_c$ is treated
in Sec.~\ref{sec:nearext}). It is asymptotically flat with a leading
$1/x^3$ tidal-charge correction $f\to1-1/x+\tfrac32(q/2)^2/x^3$, and over the
survey range its exterior Ricci scalar is \emph{negative}, decaying as
$r^{-5}$ (the horizon value turns positive for
$q>q^*=(4/5)^{3/2}\simeq0.7155$; Sec.~\ref{sec:nearext}); $q\to0$ recovers Schwarzschild.
Since the Bardeen solution -- alone among the five backgrounds -- carries
genuine matter content, we state explicitly, once for the whole paper,
that the scalar treated here is a \emph{test} field: it couples only to
the metric (and, through $\xi\mr$, to its curvature), with no direct
coupling to the nonlinear-electrodynamics field sourcing the geometry,
which therefore enters the mode equation of Sec.~\ref{sec:formalism}
only through $f$ and $\tilde g$.

\paragraph*{Scope of the calculation.} Four of the five backgrounds are not
solutions of the Einstein equations, which raises the question of which
equation the quantum field should obey on them. We adopt the standard framework
of quantum field theory on a fixed background: the scalar obeys the covariant
wave equation of Sec.~\ref{sec:formalism} on the effective geometry, and
renormalization subtracts the covariant point-splitting
counterterms~\cite{Christensen}, which are built from the local geometry
alone. No step of the calculation uses gravitational field equations, so the
computation is self-consistent for any static, spherically symmetric metric;
the same assumption underlies the quasinormal-mode, greybody and
Hawking-emission studies of these geometries mentioned in
Sec.~\ref{sec:intro}. It is, however, an assumption about the matter sector.
The KS, ZLMY and BR deformations are themselves of quantum origin (they vanish
as $\hbar\to0$), and $\phisq$ is of order $\hbar$, so the deformation imprint
computed here is formally of order $\hbar^2$ -- the same order at which a
quantum-gravitational modification of the scalar's own dynamics (for instance
polymer corrections to the matter sector in loop quantum gravity, or a running
of the scalar's couplings in the RG-improved setting) could enter. Such
modifications are model dependent, are not fixed by the effective metrics, and
are not included. Our results should therefore be read as the response of a
quantum field to the effective geometry alone, with the deformation parameters
treated phenomenologically; taken at face value, the larger deformations
correspond to near-Planckian black holes (for BR, $M/M_{\rm Pl}=
(\tilde\omega/\Omega)^{1/2}$, i.e.\ from $11$ at $\Omega=0.02$ to $3.5$ at
$\Omega_{\rm cr}$). The Bardeen geometry is not subject to these remarks: it
is a solution of general relativity with a classical source, and the
calculation is standard semiclassical gravity with a test field.

These analytic properties are used as initial numerical validation targets. 
Our symbolic engine reproduces, for every geometry, the closed-form $x_+$, $\kappa$ and
$\beta=2\pi/\kappa$ to machine precision, and its internally computed Ricci
scalar agrees with an independent symbolic evaluation to a median of
$\sim10^{-15}$. In the respective classical limits
($a_{\rm KS},\zeta,\Omega,q\to0$) the full $\vp(x)$ collapses onto the
Schwarzschild profile to better than $10^{-9}$ on a common grid -- a
same-code-path consistency test, in which the correlated
mode-sum/counterterm noise cancels in the difference; this is why the
agreement is far tighter than the $\approx3\times10^{-7}$ absolute noise
floor of Sec.~\ref{sec:results}, validating the symbolic pipeline.

\section{Formalism}
\label{sec:formalism}
We compute the renormalized coincidence limit of the Euclidean Green function for
a scalar field of mass $m$ and curvature coupling $\xi$ in the Hartle--Hawking
state at the geometry's own temperature $\beta^{-1}=\kappa/2\pi$. 

Since the computations are complex, it is useful to describe schematically the approach we adopt. We recall that the method was originally developed by Candelas \cite{Candelas:1980} and further extended by Anderson \cite{Anderson:1990jh}. Generalizations, among others, were later developed by the present author and collaborators \cite{Flachi:2008sr,Quinta:2016,Flachi:2026}. We refer
the reader to the relevant literature, which is too long to summarize here.

Consider the scalar case,
\begin{equation}
(\Box -m^2 -\xi {\cal R} ) \g (x,x') = - |g|^{-1/2}\, \delta(x - x'),
\end{equation}
with $x=(t,r,\theta,\varphi)$, ${\cal R}$ the Ricci scalar and $\xi$ a non-minimal coupling. We shall start from a sum-over-modes form of the Green function, where this is expressed as
\begin{equation}
\g(x,x') \sim \sum_{k} c_k {P^*_k(x') P_{k}(x)}.
\label{PkPk}
\end{equation}
In the above expression we have written the normal modes as $P_k(x)$, $k$ is a generic multi-index and the coefficients $c_k$ include the Wronskian of the solutions. The problem at hand is to compute the coincidence limit $x'\to x$ of an expression like the one above. This is neither trivial nor easy since the expression is divergent in the limit $x' \to x$ and the summand diverges for large $k$: neither the coincidence limit, nor the summation over $k$ can be performed in any order yielding a finite expression without renormalization. The idea of point-splitting is that as long as we keep at least one direction separate, then the coincidence limit can be taken along all other directions: in other words, if we keep the time directions separated, $t' \neq t$ -- a trick that works in general in any static background -- one can reduce an expression like (\ref{PkPk}) to a simpler form where the summand depends only on one radial coordinate:
\begin{equation}
\lim_{x' \to x} \g(x,x') \sim \lim_{t' \to t} \sum_{k} f_k(t'- t) \phi_k(r),
\end{equation}
where again we have ignored coefficients inside and outside the summation and other details irrelevant to the present schematic picture (in particular, we are assuming that the modes can be written in separable form and the angular part can be factorized inside the function $f_k$; this is always possible for free fields on a spherically symmetric background). Then, the full coincidence limit ${t' \to t}$ can be taken if counter-terms compensating for the divergences are added; these have been computed in general four-dimensional backgrounds \cite{Christensen} and allow us to obtain a regularized expression, which we formally write as
\begin{equation}
\lim_{x' \to x} \g(x,x') \sim  \sum_{k} \left[\phi_k(r)\right]_{\text{reg}},
\end{equation}
where the index \rm{reg} signifies that the counter-terms have been appropriately subtracted. The same procedure can be applied to different dimensionality, subtracting appropriate counter-terms. The remaining part of the computation requires the summation over $k$: this is problematic due to the fact that the summand diverges in the large $k$ limit\footnote{To be precise there are two types of divergences: one associated with summation over the ``energy'' eigenvalues, associated with UV divergences, and the other associated with summation over angular momentum. The latter is easily resolved, since the angular divergences are spurious. Instead, the UV divergences have to be renormalized.}. These can be dealt with by using an approximate form of the summand. A possibility is to extract the divergences by use of the WKB form of the normal modes, a procedure which yields
\begin{align}
\lim_{x' \to x} \g(x,x') \sim &\sum_{k} \left\{\left[\phi_k(r)\right]_\text{reg} - \left[\phi_k(r)\right]_\text{wkb}\right\} \nonumber\\
&+ \sum_{k}  \left[\phi_k(r)\right]_\text{wkb}.
\label{Eq:4}
\end{align}
This simple step of adding and subtracting the WKB approximants has the effect of relegating the divergences to the latter contribution. Adding the counter-terms to the above expression (see below) allows us to eliminate the divergence and arrive at a renormalized finite expression for the vacuum polarization. The usefulness of this step becomes evident once it is realized that only the first few orders in the WKB expansion are sufficient to take care of the divergences, that is, the difference between the regularized and WKB parts converges. The latter term can instead be dealt with by analytical continuation (in a form similar to dimensional or zeta-regularization) as done in Ref.~\cite{Flachi:2008sr} for the case of asymptotically AdS black holes (again, analytical continuation is a procedure that can be applied in general cases). While computing the first part is not problematic in principle, the convergence is slow: in practice, this means that the summation has to include the contribution from a large number of modes: this means that one has to solve a large number of differential equations with singular coefficients. Here, to avoid confusion, we should remark that the convergence of this summation should not be confused with the convergence of the WKB approximation. Here the WKB approximants are used solely as a way to extract the diverging behavior of the expression. Any other way to do so would be perfectly acceptable. A second remark is that in four dimensions only a finite number of WKB terms are necessary to extract the divergences and adding more WKB terms does not generally improve the convergence. One issue is then how to compute the summations efficiently, which constitutes one of the bottlenecks of the method. There are various ways to optimize this step of the computation, for example leveraging Monte Carlo-like techniques (Ref.~\cite{Flachi:2008sr} uses a simple implementation of such a procedure), however, here we are interested in a more \textit{generalist} approach requiring no approximations or ad hoc adaptations of any sort.

Concretely, the point-splitting approach allows one to write the coincidence limit of the Green function,
\begin{align}
\lim_{x'\to x}\g(x,x') &= \frac{1}{\beta}
\lim_{x'\to x}\sum_{l,n} 
e^{i\omega_n (\tau-\tau')} 
\frac{2l+1}{4 \pi} 
P_l(\cos(\gamma))\nonumber\\
&\times \frac{1}{r^2}\sqrt{\frac{1}{f \tilde g}}
\frac{q_{nl}(r_<)\, p_{nl}(r_>)}{p_{nl} q'_{nl}- q_{nl}p'_{nl} },
\end{align}
where $q$ and $p$ are the two independent solutions of the radial wave equation (regular at the horizon and decaying at infinity, respectively) and $r_>$ ($r_<$) is the larger (smaller) of the pair $\{r,r'\}$, as the following regularized expression
\begin{align}
\phisq&=\lim_{x'\to x}\Big[\,\g(x,x')-\Delta_{\rm CD}(x,x')\,\Big]\nonumber\\
      &={\alpha\over8\pi^2}\left(\Sigma_1+\Sigma_2+\Xi\right),
      \qquad \alpha\equiv{2\pi\over\beta},
\label{eq:phisq}
\end{align}
where $\Sigma=\Sigma_1+\Sigma_2$ is the mode sum over Matsubara frequencies
$\omega_n=2\pi n/\beta$ and angular momenta $l$, and $\Delta_{\rm CD}$ is the
covariant Christensen--DeWitt counterterm contribution constructed from the
local geometry. The block $\Xi$ collects, precisely, \emph{all} the remaining,
analytically computable contributions: the leading term $\Upsilon_0$, the
resummed WKB blocks (expressed as generalized $\zeta$-functions of the WKB
frequencies as in Refs.~\cite{Flachi:2008sr,Quinta:2016}), and the
counterterm quadratures arising from the coincidence limit of
$\Delta_{\rm CD}$; it is their combination that renders the result finite (see Appendix~\ref{app:code} for more details).
The terms $\Sigma_1$ and $\Sigma_2$ contain the remainder of the regularization and are written as 
\begin{equation}
\Sigma_1
=
\sum_{l=0}^{\infty}
\left(l+{1\over 2}\right)
\left(
{1\over r^{2}W_{0l}} - {1\over r^{2}\tilde W_{0l}}
\right),
\end{equation}
\begin{equation}
\Sigma_2
=
2 \sum_{n=1}^{\infty}\sum_{l=0}^{\infty}
\left(l+{1\over 2}\right)
\left(
{1\over r^{2} W_{nl}} - {1\over r^{2}\tilde W_{nl}}
\right),
\end{equation}
the function $W_{nl}$ being computed exactly from the two independent
solutions of the radial equation -- $q_{nl}$, regular at the horizon, and
$p_{nl}$, decaying at infinity -- through their logarithmic derivatives,
\begin{equation}
W_{nl}={\sqrt{f\tilde g}\over2}
\left({d\ln q_{nl}\over dr}-{d\ln p_{nl}\over dr}\right),
\label{eq:Wexact}
\end{equation}
equivalently $W_{nl}=\sqrt{f\tilde g}\,(q_{nl}'p_{nl}-q_{nl}p_{nl}')/(2\,q_{nl}p_{nl})$,
the Wronskian combination entering the mode-sum Green function above. This
is literally the combination the code evaluates: the integrator propagates
only the logarithmic derivatives $d\ln q_{nl}/dr$ and $d\ln p_{nl}/dr$
(Appendix~\ref{app:code}), so the arbitrary -- and in practice exponentially
large -- normalizations of $q_{nl}$ and $p_{nl}$ cancel identically.
The function with a tilde represents the WKB approximation counterpart of the
exact numerically computed $W$. For definiteness: the subtraction uses the
WKB solution for $W$ at next-to-leading (second) order throughout -- in four
dimensions this order suffices to extract all divergences, and higher orders
do not improve the convergence of the difference -- while the counterterms
are Christensen's covariant geodesic point-splitting expressions for
$\phisq$~\cite{Christensen}. The fully explicit general-$(f,g)$ formulae for
the WKB blocks, their $\zeta$-function resummation and the counterterm
quadratures are collected in the Appendix of Ref.~\cite{Flachi:2026} and,
for the asymptotically AdS case, in Ref.~\cite{Flachi:2008sr}, and for a Lifshitz background in Ref.~\cite{Quinta:2016}.
These differences are the numerically demanding ingredients: the exact radial
modes entering $\Sigma_1$ ($n=0$) and $\Sigma_2$ ($n\ge1$) are obtained by
integrating, for each $(n,l)$, the homogeneous radial equation by a
fourth-order Runge--Kutta scheme, starting from the regular near-horizon
expansion and imposing the asymptotically flat outer boundary condition, and
subtracting the next-to-leading WKB approximant mode by mode. Because all five
backgrounds are asymptotically flat, the outer boundary condition is the
standard one and requires no modification. The counterterms and the analytic
blocks entering $\Xi$ follow from the closed-form expressions of
Refs.~\cite{Christensen,Flachi:2008sr}, evaluated here from the symbolically
generated derivatives of $f$ and $g$. General explicit expressions can be read off Refs.~\cite{Flachi:2008sr,Quinta:2016,Flachi:2026} along with full analytic derivations. The new element is encoded in the metric functions $f$, $g$ and the
required derivatives ($f',\dots,f''''$, $g',\dots,g'''$) (which are produced
automatically for each geometry), together with the adjustment of the
convergence procedures and of the boundary conditions required by the
different backgrounds.

We emphasize that in our approach none of these ingredients assumes $g_{tt}g_{rr}=-1$: the
Christensen--DeWitt counterterms are covariant expressions evaluated on the
general line element~(\ref{eq:lineelement}) with independent $f$ and
$\tilde g$, and the radial equation and its WKB approximants carry $f$ and
$g$ (and their derivatives) separately, as the $\sqrt{f\tilde g}$ factors
above make explicit. 
In particular there
is no separate $f\neq\tilde g$ code branch: every geometry, including all
the validation cases of Sec.~\ref{sec:results} (Schwarzschild,
Reissner--Nordstr\"om, and the classical limits), runs through the identical
general-$(f,g)$ expressions, which the $f=\tilde g$ cases probe on that
subspace; off the subspace, the engine's internally computed Ricci scalar
agrees with an independent symbolic evaluation for ZLMY-2 at the same
$\sim10^{-15}$ level as for the other geometries
(Sec.~\ref{sec:geometries}), directly checking the curvature and
counterterm inputs, and a quantitative scaling check of the $f\neq\tilde g$
response is presented in Sec.~\ref{sec:results}.

\section{Implementation and high-performance computation}
\label{sec:implementation}
The computational core of the project is a Fortran implementation of the point-splitting
scheme originally developed for the test computation of the vacuum polarization on the loop-quantum-gravity geometry in Ref.~\cite{Flachi:2026} and extended for the classes of solutions considered in this work. 
The code evaluates the renormalized coincidence limit
as WKB-subtracted mode sums (a static angular sum computed to quadruple
precision and a Matsubara double sum over $(n,\ell)$), together
with the counterterm quadratures, on a fixed radial grid of $10^4$
logarithmically spaced points. The grid, and every mode integration on it, spans
$x\in[x_++10^{-4},\,200]$ with the innermost region $x-x_+<10^{-2}$ excised
from all reported profiles due to boundary-data
transient contamination: the truncated Frobenius/Eddington--Finkelstein expansions that
seed each mode at the first grid point $x_++\varepsilon$,
$\varepsilon=10^{-4}$ (Appendix~\ref{app:code}), carry a small admixture of
the discarded irregular branch that dies off only over a finite range of
$x-x_+$, and the fixed $(n,\ell)$ truncations of the mode sums lose
accuracy as the horizon blueshift grows. In practice the transient is
steep -- it falls from $O(10^3)$ in $\vp$ at the first grid points to below
$10^{-5}$ by $x-x_+\simeq4\times10^{-3}$ -- so the excision at $10^{-2}$,
two decades above $\varepsilon$, is deliberately conservative, keeping a
buffer between any residual transient and the extrapolation window;
resolving the excised zone instead would require jointly refining
$\varepsilon$ and the mode truncations for a marginal gain in accuracy.
Results are therefore quoted on $x\in[x_++10^{-2},\,200]$ and horizon
values are obtained by the calibrated extrapolation of
Sec.~\ref{sec:results}. The radial mode functions are integrated by a fourth-order Runge--Kutta scheme
with boundary conditions from a regular-horizon expansion on the inner side
and the WKB asymptotics on the outer side. Because the renormalized
$\phisq$ emerges from cancellations spanning several orders of magnitude
between the mode sums and the counterterms (the intrinsic noise floor is
$\approx3\times10^{-7}$ in $\vp$; Sec.~\ref{sec:results}), the code is compiled with strict IEEE floating-point
semantics only (i.e., all results were obtained with strict IEEE 754 arithmetic; no fast-math or value-unsafe optimizations were enabled).

For the present work this core was extended in four respects:\\ 
(\textit{i})~A \emph{metric-agnostic} front end: the lapse and radial functions are supplied
symbolically in a configuration file, and all derivatives entering the mode
equations and counterterms (up to fourth order) are generated analytically by
a computer-algebra layer and spliced into the Fortran source, so that no
numerical differentiation of the metric functions, curvature terms or
counterterm inputs occurs anywhere -- the single, deliberate exception, a
finite-difference Taylor seed for the static near-horizon boundary
condition, is described in Appendix~\ref{app:modes} -- and a single
validated code base serves every geometry; the horizon location and surface gravity are
determined at startup by a dense logarithmic scan with bisection and Newton
refinement, robust also for the nearly degenerate double horizons of the
near-extremal configurations.\\ 
(\textit{ii})~
Runge--Kutta stage coefficients precomputed on the fixed grid,
geometry factors hoisted out of the innermost quadrature loops\footnote{Geometry factors (Jacobians, metric terms, radii, quantities encoding the mesh/coordinate shape) do not depend on the innermost loop index, so evaluating them there is wasted work repeated millions of times. ``Hoisting''", i.e. moving a loop-invariant computation outside the loop, speeds up the computation.}, and OpenMP
parallelism over mode and component sums -- yielding an order-of-magnitude
single-node speedup, to $\sim\!70$--$90$~s per case at $16$ threads;
this is what makes sweeps of $10^2$--$10^3$ configurations practical.\\ 
(\textit{iii})~The angular-truncation selection of the thermal sum was made a smooth regularized
function of radius, 
eliminating small piecewise-constant discretization offsets from the
profiles; the raw output then reproduces the analytic Schwarzschild
DeWitt--Schwinger tail with no post-processing (Sec.~\ref{sec:results}).\\ 
(\textit{iv})~Every run goes through automated checks (divergence, non-finite output, fallback conditions) and
the extended code was regression-tested against the reference implementation of Ref.~\cite{Flachi:2026},
agreeing to within $10^{-12}$ of the local $\vp$ scale.
A description of the numerical engine detailed enough to allow an
independent reconstruction, including the design choices responsible for
its speed, is given in Appendix~\ref{app:code}.

The production runs were executed on the Miyabi Supercomputing System
(Joint Center for Advanced High Performance Computing -- JCAHPC). Each case is a single-node
shared-memory computation, and sweeps are parallelized across
configurations: cases were packed six-to-seven at a time, at $16$--$18$
OpenMP threads each, onto Miyabi-C nodes (dual Intel Xeon Platinum 8480+,
$112$ cores per node) through the Portable Batch System (PBS). The unified data set
underlying Secs.~\ref{sec:results} and~\ref{sec:nearext} -- $430$
configurations: a main grid of $360$ spanning seven metrics, five field masses
($\mu=0.2,0.5,1.0$ plus the light validation masses $\mu=0.025,0.05$) and two
curvature couplings, plus the light-mass anchors, the near-extremal scans and
the $\xi$-scans -- completed in about $1.5$~h of wall time on two nodes,
producing $\sim\!11$~GB of raw
output. 
The deepest near-extremal cases whose
results are used in Secs.~\ref{sec:results} and~\ref{sec:nearext} reach
$\beta/r_s\simeq2.2\times10^3$; the coldest configurations attempted (Reissner--Nordstr\"om convergence probes at $\beta/r_s\simeq7\times10^3$, with Matsubara cutoffs of order $2\times10^4$, which lie beyond the
method-validity boundary established in Sec.~\ref{sec:results} and are
therefore excluded from the physics analysis) still run in under an hour
each. Bitwise-class reproducibility was
verified between the development workstation and Miyabi (identical source code
and flags (compiler options); maximum profile deviation $2\times10^{-12}$), a nontrivial check
given the cancellation-dominated arithmetic.

All results presented in this paper have been uploaded on Zenodo~\cite{zenodo_repository}.

\section{Numerical analysis and results}
\label{sec:results}

\begin{figure*}[t]
\centering
\includegraphics[width=\textwidth]{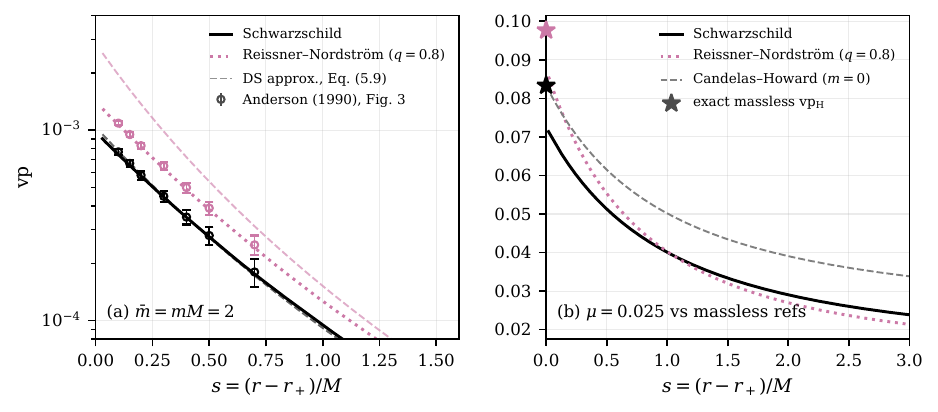}
\caption{Validation against published results, in Anderson's radial variable
$s=(r-r_+)/M$. Colour identifies the metric in both panels: black for
Schwarzschild, pink for Reissner--Nordstr\"om at $q=0.8$.
(a)~$\bar m=mM=2$: our raw computed profiles (solid:
Schwarzschild; dotted: RN) versus the
numerical results digitized from Fig.~3 of Ref.~\cite{Anderson:1990jh}
(open circles, error bars = digitization uncertainty) and his large-mass
DeWitt--Schwinger approximation [Eq.~(5.9) there, thin dashed, same
colour as the corresponding metric].
(b)~near-massless field ($\mu=0.025$): our profiles versus the exact massless
horizon values (stars, coloured by metric; Candelas's $1/12$ for Schwarzschild and Frolov's
$\tfrac23 s_c/(1+s_c)^3$~\cite{Frolov:1982}, $s_c=\sqrt{1-q^2}$, for $q=0.8$)
and the Candelas--Howard massless Schwarzschild profile (gray dashed); the
small offset is the expected $(1-3mr_+)$ mass correction.}
\label{fig:anderson}
\end{figure*}

A convention used in all figures: symbols mark individually computed
configurations (one complete mode-sum computation each); the thin, lighter
lines through them are monotone cubic (PCHIP) interpolations, drawn only as
guides to the eye to identify each series, and carry no information beyond the
computed points. Continuous curves without symbols are computed radial
profiles, plotted on the full $10^4$-point grid of each run; analytic
references, approximations and fits are identified as such in each caption.
It is useful to state the uncertainty budget of the calculation. Because $\phisq$ arises as a near-cancellation between the large
mode sum and the large covariant counterterms (each of order unity), the
result carries an intrinsic \emph{absolute} noise floor, measured in the far
zone where the physical signal has decayed and identical in the
Schwarzschild reference: $\approx3\times10^{-7}$ in $\vp$
(Fig.~\ref{fig:dstail}).\footnote{In plain terms, the answer is a tiny
difference of two large numbers: the $\sim\!10^{-7}$ numerical error of the
individual pieces does not cancel in the subtraction, and its impact is
therefore \emph{relative to the local signal} -- a few parts in $10^{5}$ on
horizon values $\vp_{\rm H}\sim10^{-2}$, of order $10^{-3}$ in the mid zone
where $\vp\sim\text{few}\times10^{-4}$, and of order unity beyond
$r/r_s\simeq4$, where the profiles are noise-dominated.} On horizon values
the total uncertainty is instead dominated by the extrapolation and by
method-level systematics, quantified below: the window stability is
$0.2\%$ (reaching $0.6\%$ at the largest KS deformation), the cross-method
comparison with Anderson's approach indicates an
absolute-scale systematic of $\lesssim0.5\%$, and differences or ratios
between geometries -- produced by a single code path with identical grid,
boundary treatment and extrapolation window -- are reliable at the
window-stability level. 
The cross-method calibration rests entirely on $\mr=0$
references (Schwarzschild and RN, the only backgrounds with published
benchmarks), whereas the five new geometries have $\mr\neq0$, which is
precisely where the curvature content of the counterterms is exercised;
that sector is validated instead by the classical-limit collapse
(Sec.~\ref{sec:geometries}), by the parameter-free far-zone $[a_2]$
comparison, and by the horizon $\xi$-slope tests of
Table~\ref{tab:xislope}, all of which probe the curvature terms with
essentially no adjustable parameters. A cross-method computation of a
deformed configuration would be an optimal choice but it is beyond the
present scope. The four digits quoted in
Tables~\ref{tab:vpH_param}--\ref{tab:vpH_Bardeen} serve this internal
comparability; absolute values carry the $\lesssim0.5\%$ systematic. The
smallest tabulated effect (ZLMY-2 at $\zeta=0.1$, $+0.3\%$) is at the edge
of significance as an isolated absolute statement, but is established by
the smooth $\zeta^2$ trend of the full grid discussed below.
As in the earlier study of Ref.~\cite{Flachi:2026}, the angular sum in $\Sigma_2$ is
truncated at the grid-independent plateau of its cumulative value, and we monitor
convergence under grid refinement ($N=10^4$ throughout, cross-checked against
finer grids). The horizon value $\vp_{\rm H}$ is obtained by a quadratic
extrapolation of the near-horizon profile over the window
$x-x_+\in[1.0,\,2.5]\times10^{-2}$, chosen to lie entirely outside the
boundary-data transient zone that contaminates the mode sum
for $x-x_+\lesssim10^{-2}$ (Sec.~\ref{sec:implementation}). The windows
tested are: lower edge varied over $[1.0,1.5]\times10^{-2}$, upper edge
over $[2.0,3.5]\times10^{-2}$, fit order quadratic and cubic. Under these
variations the extrapolated value shifts by $0.1$--$0.3\%$ across the main
survey, reaching $0.6\%$ in the worst case (the largest KS deformation
with the window shifted outward; a few percent for the most strongly
curved near-extremal profiles of Sec.~\ref{sec:nearext}), while a cubic
fit on the widest window agrees with the baseline to $\lesssim0.1\%$.
\emph{Differences} between geometries cancel this window systematic
almost entirely: the ZLMY-2 excess over Schwarzschild at $\zeta=0.1$
($+0.325\%$, the smallest effect we quote) moves by less than
$0.002$ percentage points across all window variants. For Schwarzschild at $\mu=1/2$ the extrapolation gives
$\vp_{\rm H}=1.074\times10^{-2}$, consistent within half a percent with
the value $1.069\times10^{-2}$ obtained by Anderson~\cite{Anderson:1990jh}. We validated the pipeline directly against the published
Reissner--Nordstr\"om results of Anderson~\cite{Anderson:1990jh}
(Fig.~\ref{fig:anderson}): at $\bar m=mM=2$ our raw profiles pass through the
numerical results of his Fig.~3 (digitized, $\pm3\times10^{-5}$) to
within $1$--$2\%$ for both Schwarzschild and Reissner--Nordstr\"om with charge $q\equiv Q/M=0.8$, while his large-mass
DeWitt--Schwinger formula overestimates near the horizon exactly as he notes;
in the near-massless limit ($\mu=0.025$) our horizon values sit the expected
small-mass factor $(1-3mr_+)$ below the exact massless values [Candelas's
$1/12$~\cite{Candelas:1980} and Frolov's $\tfrac23 s_c/(1+s_c)^3$,
$s_c=\sqrt{1-q^2}$~\cite{Frolov:1982}], and the profile tracks
the Candelas--Howard massless curve~\cite{CandelasHoward}. Each
geometry was run over a grid of its deformation parameter, three field masses
($\mu=0.2,0.5,1.0$) and two couplings ($\xi=0,1/6$); the data set is uniform --
every case is produced by the same code path, radial grid and convergence
protocol -- and machine-generated, in that the geometry-specific source and the
parameter sweep are built automatically from the symbolic metrics
(Sec.~\ref{sec:implementation}), with no per-case tuning.

\begin{figure*}[t]
\centering
\includegraphics[width=\textwidth]{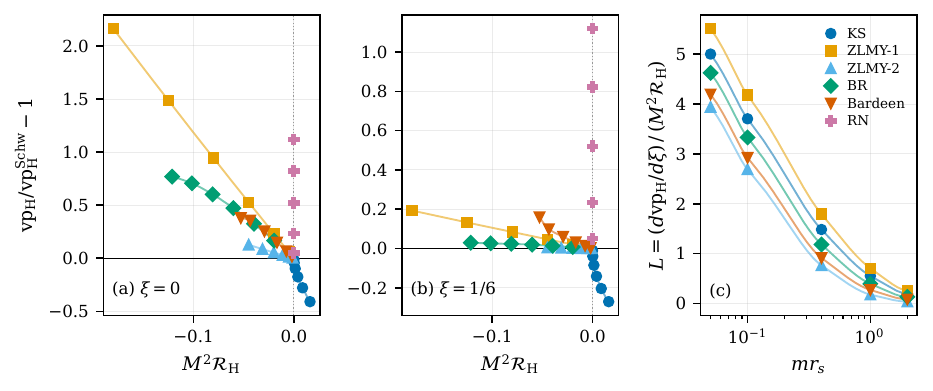}
\caption{Horizon response of all geometries at $\mu=1/2$ versus the
horizon Ricci scalar $M^2\mr_{\rm H}$, a measure of the deformation with the
same meaning in every theory. (a)~$\xi=0$ and (b)~$\xi=1/6$: relative change of
$\vp_{\rm H}$ with respect to Schwarzschild over the full deformation grids
(KS $a/r_s=0.1$--$0.6$; ZLMY $\zeta/r_s=0.1$--$0.6$; BR $\Omega=0.02$--$0.12$;
Bardeen $q=0.1$--$0.6$) and for Reissner--Nordstr\"om ($q=0.3$--$0.95$), which
has $\mr=0$ and lies on the vertical axis. (c)~$\xi$-response per unit horizon
curvature, $L=(d\vp_{\rm H}/d\xi)/(M^2\mr_{\rm H})$, from the $(\xi=0,\tfrac16)$
pairs, versus field mass, at the largest deformation of each geometry.
Symbols: computed configurations; thin lines: guide-to-the-eye
interpolations.}
\label{fig:overlay_vpH}
\end{figure*}

\begin{figure}[t]
\centering
\includegraphics[width=\columnwidth]{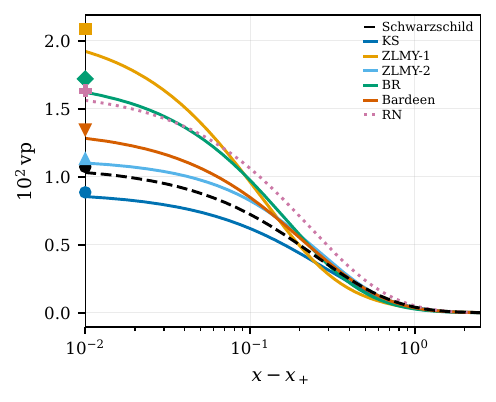}
\caption{Radial profiles $\vp(x)$ versus the distance from the horizon
$x-x_+$ (log scale) at $\mu=1/2$, $\xi=0$, one representative deformation per
geometry (KS $a/r_s=0.4$; ZLMY $\zeta/r_s=0.4$; BR $\Omega=0.08$; Bardeen
$q=0.4$; RN $q=0.8$), with the Schwarzschild profile dashed. The curves
are the computed profiles on the full $10^4$-point radial grid. Symbols at the left
edge mark the extrapolated horizon values $\vp_{\rm H}$: every profile
decreases monotonically from its horizon. (The innermost
$x-x_+<10^{-2}$ zone, where the boundary-data transient of
Sec.~\ref{sec:implementation} contaminates the mode sum, is excluded; horizon values come from the
calibrated extrapolation.) All profiles decay to the common massive-field
exterior tail.}
\label{fig:overlay_profiles}
\end{figure}

\begin{figure}[t]
\centering
\includegraphics[width=\columnwidth]{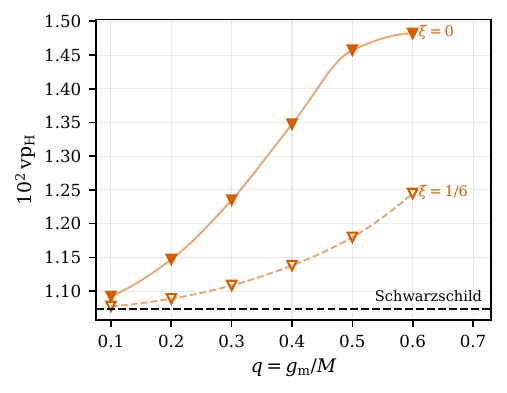}
\caption{Bardeen horizon vacuum polarization $\vp_{\rm H}$ versus the charge
$q=g_{\rm m}/M$ at $\mu=1/2$, for minimal ($\xi=0$) and conformal ($\xi=1/6$) coupling.
The deformation \emph{enhances} $\vp_{\rm H}$ (negative exterior Ricci scalar);
conformal coupling flattens the trend, the signature of the DeWitt--Schwinger
term linear in $(\xi-\tfrac16)\mr$ (the residual $\xi=1/6$ trend is analyzed
in the text). The black dashed line is the Schwarzschild value. Filled
(open) symbols: computed configurations at $\xi=0$ ($\xi=1/6$); thin solid (dashed) lines: guide-to-the-eye interpolations.}
\label{fig:bardeen_trends}
\end{figure}

\begin{figure}[t]
\centering
\includegraphics[width=\columnwidth]{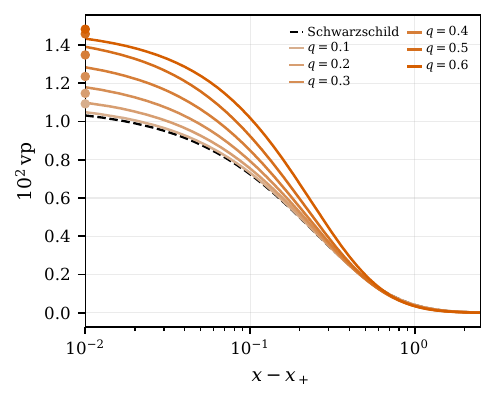}
\caption{Bardeen radial profiles $\vp$ versus the distance from the
($q$-dependent) outer horizon $x-x_+$ (log scale) at $\mu=1/2$, $\xi=0$ across
the charge grid $q=0.1$--$0.6$ (light to dark; Schwarzschild dashed),
computed on the full radial grid. Each
curve peaks at its horizon and decays monotonically through the near zone.}
\label{fig:bardeen_profiles}
\end{figure}

\begin{figure}[t]
\centering
\includegraphics[width=\columnwidth]{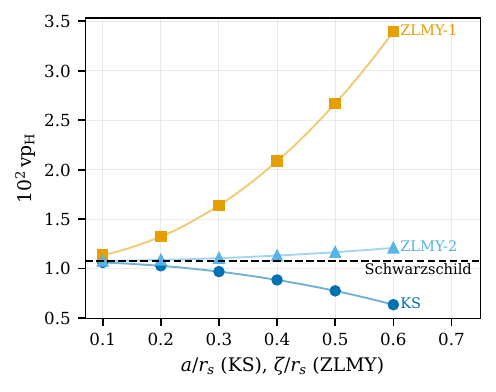}
\caption{Horizon vacuum polarization $\vp_{\rm H}$ versus the quantum-deformation
parameter ($a_{\rm KS}$ for KS, $\zeta$ for ZLMY) at $\mu=1/2$, $\xi=0$. KS
suppresses, ZLMY-1 and ZLMY-2 enhance; the dashed line is
the Schwarzschild value. Symbols: computed configurations; thin lines: guide-to-the-eye interpolations.}
\label{fig:trends}
\end{figure}

\begin{figure}[t]
\centering
\includegraphics[width=\columnwidth]{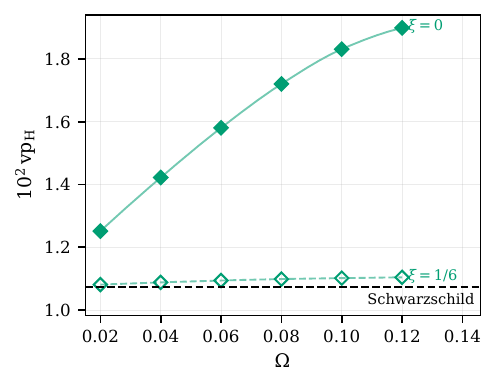}
\caption{Bonanno--Reuter horizon polarization versus $\Omega$ at $\mu=1/2$, for
minimal and conformal coupling. The minimal-coupling curve rises by $\sim77\%$
toward the extremal value; conformal coupling collapses the trend onto the
Schwarzschild line, the signature of the DeWitt--Schwinger curvature term
linear in $(\xi-\tfrac16)\mr$. Filled (open) symbols: computed
configurations at $\xi=0$ ($\xi=1/6$); thin solid (dashed) lines: guide-to-the-eye interpolations; black dashed line:
Schwarzschild.}
\label{fig:trends_BR}
\end{figure}

\paragraph*{Horizon polarization.}
Table~\ref{tab:vpH_param} collects the horizon polarization of the three
$\zeta$/$a$-type geometries at $\mu=1/2$, $\xi=0$, and Table~\ref{tab:vpH_BR} that
of BR. Within each geometry, the response over the swept range is:
\begin{itemize}
\item \textbf{KS suppresses} $\vp_{\rm H}$, by $\sim40\%$ from
  $a_{\rm KS}=0.1$ to $0.6$ -- consistent with its \emph{positive} exterior Ricci
  scalar;
\item \textbf{ZLMY-1 enhances} by a factor $\sim3$ over the same range;
\item \textbf{ZLMY-2 enhances}, by $\sim13\%$ -- weaker than ZLMY-1 even at
  equal horizon curvature (Fig.~\ref{fig:overlay_vpH}), the geometry that deforms only $g_{rr}$;
\item \textbf{BR enhances}, the excess over Schwarzschild growing to $\sim77\%$ from
  $\Omega=0.02$ to $0.12$, with the curve bending over as the extremal value is
  approached;
\item \textbf{Bardeen enhances}, $\vp_{\rm H}$ growing by $\sim38\%$
  from $q=0$ to $q=0.6$ at $\mu=1/2$, $\xi=0$
  (Table~\ref{tab:vpH_Bardeen}) -- consistent with its
  \emph{negative} exterior Ricci scalar.
\end{itemize}
Figures~\ref{fig:trends},~\ref{fig:trends_BR} and~\ref{fig:bardeen_trends}
display these trends; Fig.~\ref{fig:bardeen_profiles} shows the Bardeen radial
profiles, and Fig.~\ref{fig:overlay_profiles} overlays one representative radial
profile per geometry. These percentages depend on the swept ranges, which
are not matched between theories in any physical sense, and are not meant to
be compared between geometries.

A comparison across theories requires a measure of the deformation with
the same physical meaning in each; the parameters $a$, $\zeta$, $\Omega$ and
$q$ do not provide one. We use the horizon Ricci scalar, which controls the
response directly through Eq.~(\ref{eq:slopepred}), and collect all
geometries against $M^2\mr_{\rm H}$ in Fig.~\ref{fig:overlay_vpH}.
(a)~At $\xi=0$, ZLMY-1, BR and Bardeen fall on a common band: at
$M^2\mr_{\rm H}=-0.05$ the relative enhancement is $0.59$, $0.40$ and $0.38$,
respectively, i.e.\ responses comparable to within a factor $1.6$ at equal
curvature. ZLMY-2, whose lapse is undeformed, responds about four times more
weakly at equal curvature. KS ($\mr_{\rm H}>0$, $M^2\mr_{\rm H}\le0.016$)
and Reissner--Nordstr\"om ($\mr=0$) show that the total response is not a
function of $\mr_{\rm H}$ alone. (b)~At $\xi=1/6$, where the term linear in
$(\xi-\tfrac16)\mr$ vanishes, a geometry-specific residual remains -- from
$-27\%$ (KS) to $+19\%$ (ZLMY-1) at the largest deformations, and up to
$+112\%$ for Reissner--Nordstr\"om -- reflecting the dependence on the horizon
data $(\kappa,r_+)$ discussed below. (c)~The $\xi$-dependent part per unit
horizon curvature, $L=(d\vp_{\rm H}/d\xi)/(M^2\mr_{\rm H})$, is common to
the five geometries to within $\pm20\%$ at light mass ($L=3.9$--$5.5$ at
$mr_s=0.05$, largest deformations); its residual spread is of the size of the
temperature dependence of the logarithm in Eq.~(\ref{eq:a1finite}), BR and
Bardeen being $\approx22\%$ colder than KS and ZLMY at these deformations.
Cross-geometry statements in this paper refer to these physical measures.

The ZLMY-2 grid deserves a dedicated comment, since it is the one geometry
that exercises the independent-$(f,\tilde g)$ sector of the machinery
(Sec.~\ref{sec:formalism}) while also carrying the weakest signal. Because
its deformation enters the radial function alone, with amplitude $\zeta^2$
[the lapse is exactly Schwarzschild], linear response requires the horizon
excess to scale as $\zeta^2$; the data do precisely this, with
$\vp_{\rm H}/\vp_{\rm H}^{\rm Schw}-1 = C\,\zeta^2$ and a coefficient
drifting smoothly from $C=0.326$ at $\zeta=0.1$ to $C=0.353$ at $\zeta=0.6$
-- constant to $8\%$ over a $36$-fold range of the deformation amplitude,
and connecting continuously to the validated Schwarzschild point at
$\zeta=0$. Together with the off-subspace curvature check of
Sec.~\ref{sec:formalism} and the far-zone DeWitt--Schwinger comparison
below, this is a direct functional test of the $f\neq\tilde g$ code path:
an error specific to the independent-$\tilde g$ sector would break the pure
$\zeta^2$ scaling, not merely rescale it.

\begin{table}[t]
\centering
\caption{Horizon vacuum polarization $\vp_{\rm H}=16\pi^2M^2\phisq_{\rm H}$ for
the KS, ZLMY-1 and ZLMY-2 geometries as a function of the quantum-deformation
parameter ($a_{\rm KS}$ or $\zeta$), at $\mu=1/2$, $\xi=0$. In the classical
limit (parameter $\to0$) every column reduces to the Schwarzschild
reference $\vp_{\rm H}=1.074\times10^{-2}$. Here and in
Tables~\ref{tab:vpH_BR} and~\ref{tab:vpH_Bardeen}, relative trends within a
column are stable at the $\sim0.2\%$ level, while absolute values carry a
$\lesssim0.5\%$ method-level systematic (see text).}
\label{tab:vpH_param}
\begin{ruledtabular}
\begin{tabular}{cccc}
param & KS & ZLMY-1 & ZLMY-2 \\
\hline
0.1 & $1.062\!\times\!10^{-2}$ & $1.136\!\times\!10^{-2}$ & $1.077\!\times\!10^{-2}$\\
0.2 & $1.028\!\times\!10^{-2}$ & $1.324\!\times\!10^{-2}$ & $1.088\!\times\!10^{-2}$\\
0.3 & $9.698\!\times\!10^{-3}$ & $1.639\!\times\!10^{-2}$ & $1.106\!\times\!10^{-2}$\\
0.4 & $8.858\!\times\!10^{-3}$ & $2.086\!\times\!10^{-2}$ & $1.131\!\times\!10^{-2}$\\
0.5 & $7.751\!\times\!10^{-3}$ & $2.669\!\times\!10^{-2}$ & $1.166\!\times\!10^{-2}$\\
0.6 & $6.353\!\times\!10^{-3}$ & $3.395\!\times\!10^{-2}$ & $1.210\!\times\!10^{-2}$\\
\end{tabular}
\end{ruledtabular}
\end{table}

\begin{table}[t]
\centering
\caption{Horizon vacuum polarization for the Bonanno--Reuter geometry
($\gamma=9/2$) versus $\Omega$, at $\mu=1/2$, for minimal ($\xi=0$) and conformal
($\xi=1/6$) coupling. The extremal value is $\Omega_{\rm cr}\simeq0.204$.}
\label{tab:vpH_BR}
\begin{ruledtabular}
\begin{tabular}{ccc}
$\Omega$ & $\xi=0$ & $\xi=1/6$ \\
\hline
0.02 & $1.252\!\times\!10^{-2}$ & $1.081\!\times\!10^{-2}$\\
0.04 & $1.422\!\times\!10^{-2}$ & $1.088\!\times\!10^{-2}$\\
0.06 & $1.581\!\times\!10^{-2}$ & $1.094\!\times\!10^{-2}$\\
0.08 & $1.720\!\times\!10^{-2}$ & $1.099\!\times\!10^{-2}$\\
0.10 & $1.831\!\times\!10^{-2}$ & $1.102\!\times\!10^{-2}$\\
0.12 & $1.899\!\times\!10^{-2}$ & $1.104\!\times\!10^{-2}$\\
\end{tabular}
\end{ruledtabular}
\end{table}

\begin{table}[t]
\centering
\caption{Horizon vacuum polarization for the Bardeen regular black hole versus
the charge $q=g_{\rm m}/M$, at $\mu=1/2$, for minimal ($\xi=0$) and conformal
($\xi=1/6$) coupling. The $q\to0$ value is the Schwarzschild reference
($1.074\times10^{-2}$); the extremal charge is $q_c\simeq0.770$.}
\label{tab:vpH_Bardeen}
\begin{ruledtabular}
\begin{tabular}{ccc}
$q$ & $\xi=0$ & $\xi=1/6$ \\
\hline
0.1 & $1.092\!\times\!10^{-2}$ & $1.077\!\times\!10^{-2}$\\
0.2 & $1.147\!\times\!10^{-2}$ & $1.089\!\times\!10^{-2}$\\
0.3 & $1.235\!\times\!10^{-2}$ & $1.108\!\times\!10^{-2}$\\
0.4 & $1.347\!\times\!10^{-2}$ & $1.138\!\times\!10^{-2}$\\
0.5 & $1.457\!\times\!10^{-2}$ & $1.180\!\times\!10^{-2}$\\
0.6 & $1.482\!\times\!10^{-2}$ & $1.245\!\times\!10^{-2}$\\
\end{tabular}
\end{ruledtabular}
\end{table}

\paragraph*{Mass and coupling dependence.}
The dependence on field mass and curvature coupling is uniform across geometries
and mirrors the Schwarzschild expectation: a heavier field suppresses $\phisq$
(massive fluctuations decouple), while conformal coupling moves it toward the
Schwarzschild value -- lowering it relative to the minimal case for the
negative-curvature geometries and raising it for KS. The BR data make the latter especially transparent
(Table~\ref{tab:vpH_BR} and Fig.~\ref{fig:massxi_BR}): at $\xi=1/6$ the entire
$\Omega$-trend flattens onto the Schwarzschild value (the ratio to $\xi=0$ falls
from $0.86$ at $\Omega=0.02$ to $0.58$ at $\Omega=0.12$). The Bardeen grid
behaves similarly (Fig.~\ref{fig:massxi_Bardeen}): conformal coupling reduces
the enhancement over $q=0$--$0.6$ from $+38\%$ to $+16\%$. This is the
signature that the deformation response is the field's DeWitt--Schwinger
reaction to the non-zero Ricci scalar. The structure at work near the horizon is the term \emph{linear} in the
coupling, $\propto(\xi-\tfrac16)\mr$, and it can be exhibited explicitly.
Our conventions are those of Eq.~(\ref{eq:a2}): signature $({-}{+}{+}{+})$,
curvature such that $\mr>0$ in de~Sitter space, $\sigma$ one half of the
squared geodesic separation, $\gamma_E$ the Euler constant. At first order
in the heat-kernel expansion the DeWitt--Schwinger Green function contains
the block $\Delta^{1/2}[a_1]\,K_0(m\sqrt{2\sigma})/(8\pi^2)$, with
$[a_1]=(\tfrac16-\xi)\mr$ the first curvature coefficient, so the
counterterm $\Delta_{\rm CD}$ of Eq.~(\ref{eq:phisq}) subtracts -- besides
the $\xi$-independent $1/\sigma$ and $m^2\ln\sigma$ structures -- the
logarithm
\begin{equation}
\Delta_{\rm CD}\;\supset\;
\frac{(\xi-\tfrac16)\,\mr}{16\pi^2}
\left[\,\ln\frac{m^2\sigma}{2}+2\gamma_E\right].
\label{eq:a1ct}
\end{equation}
In the thermal state the mode sum
generates the same logarithm with $\sigma$ saturated by the thermal
wavelength rather than by the point separation -- in flat space, exactly,
$\phisq_T=\tfrac{T^2}{12}-\tfrac{mT}{4\pi}
-\tfrac{m^2}{8\pi^2}\big[\ln\tfrac{m\,e^{\gamma_E}}{4\pi T}-\tfrac12\big]
+\dots$~\cite{DolanJackiw,HaberWeldon}, and in a curved static background the curvature inherits this
logarithm through the heat-kernel combination
$m^2\to m^2+(\xi-\tfrac16)\mr$~\cite{DowkerKennedy} -- so a \emph{finite}
remainder linear in $(\xi-\tfrac16)\mr$ survives the subtraction,
\begin{equation}
\delta\phisq=\frac{(\xi-\tfrac16)\,\mr}{16\pi^2}\,L,
\qquad
L=2\ln\frac{4\pi T}{m\,e^{\gamma_E}}+c\,,
\label{eq:a1finite}
\end{equation}
with $c$ an $O(1)$ constant that the leading-logarithm argument does not
fix (naive flat-space matching gives $c=1$); the data below show that at
the horizon the saturation temperature is the geometry's own Hawking
temperature. Equation~(\ref{eq:a1finite}) presupposes that the thermal
wavelength sets the infrared scale of the logarithm, i.e.\ that its argument
$4\pi T_{\rm H}/(m\,e^{\gamma_E})$ exceeds unity; for $T_{\rm H}\ll m$, as in
the near-extremal regime of Sec.~\ref{sec:nearext}, the scale is set by the
geometry instead and Eq.~(\ref{eq:a1finite}) does not apply. For the horizon value, Eq.~(\ref{eq:a1finite}) predicts the
slope
\begin{equation}
\frac{d\,\vp_{\rm H}}{d\xi}\;=\;M^2\,\mr_{\rm H}\,L\,,
\label{eq:slopepred}
\end{equation}
with $\mr_{\rm H}$ the horizon Ricci scalar and $L>0$ for light fields:
suppression at minimal coupling for $\mr>0$ (KS), enhancement for $\mr<0$
(ZLMY, BR, Bardeen), both switching off linearly at $\xi=\tfrac16$, and a
vanishing slope for the $\mr=0$ references (Schwarzschild and RN are
$\xi$-independent in our runs to $\sim10^{-12}$). For heavy fields
($mr_+\gtrsim1$) the logarithm closes and the $\xi$-linear slope crosses
over to the local large-mass form obtained from Eq.~(\ref{eq:a2}),
\begin{equation}
\frac{d\,\vp}{d\xi}\;=\;\frac{M^2}{m^2}
\left[(\xi-\tfrac16)\,\mr^2-\tfrac16\,\Box\mr\right],
\label{eq:slopea2}
\end{equation}
dominated at the horizon by the $\Box\mr$ term [the quadratic
$(\xi-\tfrac16)^2$ term of $[a_2]$ is negligible there]. The sign of the
response tracks the sign of $\mr$ -- suppression for KS ($\mr>0$),
enhancement for the negative-$\mr$ geometries (ZLMY, BR, Bardeen) --
consistent with the curvature-response interpretation established for the
loop-quantum-gravity case of Ref.~\cite{Flachi:2026}; a quantitative test of
Eqs.~(\ref{eq:slopepred})--(\ref{eq:slopea2}) against the measured slopes
is given below (Table~\ref{tab:xislope}).

The flattening at conformal coupling is strong but not complete: at
$\xi=1/6$ the residual trends at maximal deformation are $+2.9\%$ (BR),
$+0.4\%$ (ZLMY-2), $+19\%$ (ZLMY-1), $+16\%$ (Bardeen) and $-27\%$ (KS,
down from $-41\%$). These residuals collect the contributions that do
\emph{not} switch off with the linear term: the $\xi$-independent curvature
structures of the DeWitt--Schwinger expansion [at $\xi=1/6$ the
coefficient~(\ref{eq:a2}) reduces to
$\tfrac1{180}(\Box\mr+\mr_{\mu\nu\rho\sigma}\mr^{\mu\nu\rho\sigma}
-\mr_{\mu\nu}\mr^{\mu\nu})$, which does not vanish], together with the
purely metric dependence of the thermal state on the horizon data
$(\kappa,r_+)$, which drift with the deformation for KS, BR and Bardeen. The
second contribution is calibrated by Reissner--Nordstr\"om ($\mr=0$), for
which $\vp_{\rm H}$ is exactly $\xi$-independent in our runs (the $\xi=0$
and $\xi=1/6$ values agree to $\sim10^{-12}$) and yet sits $52\%$ above
Schwarzschild at $q=0.8$: a horizon-data shift of this purely classical
type persists at conformal coupling wherever the deformation moves
$(\kappa,r_+)$.

\begin{figure}[t]
\centering
\includegraphics[width=\columnwidth]{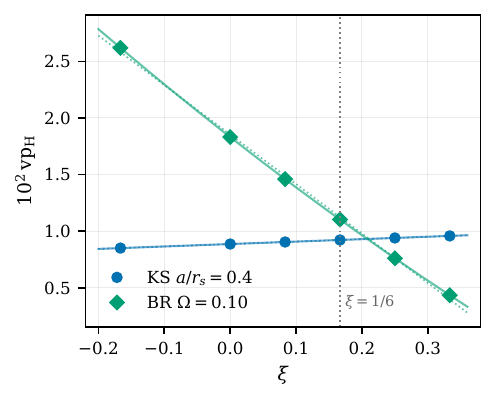}
\caption{Horizon polarization versus curvature coupling $\xi$ at $\mu=1/2$,
for KS ($a/r_s=0.4$, $\mr>0$) and BR ($\Omega=0.10$, $\mr<0$). Symbols:
computed configurations. Solid: quadratic
fit; dotted: linear fit (the two are indistinguishable). The response is linear
in $(\xi-\tfrac16)$ with the slope sign set by the sign of the Ricci scalar.}
\label{fig:xiscan}
\end{figure}

We probed the $\xi$-dependence directly with a six-point scan
$\xi\in\{-1/6,\,0,\,1/12,\,1/6,\,1/4,\,1/3\}$ at $\mu=1/2$ for one geometry of
each Ricci sign (KS at $a/r_s=0.4$; BR at $\Omega=0.10$),
Fig.~\ref{fig:xiscan}. The horizon response is \emph{linear} in $\xi$ to
excellent accuracy over the full range -- rising for KS ($\mr>0$), falling
steeply for BR ($\mr<0$), with slopes of opposite sign as dictated by the
DeWitt--Schwinger coupling term linear in $(\xi-\tfrac16)\mr$ -- and shows
no interior extremum: the quadratic $(\xi-\tfrac16)^2$ contribution is
subleading at the horizon for these backgrounds.

The measured slopes can now be compared quantitatively with
Eqs.~(\ref{eq:slopepred})--(\ref{eq:slopea2}). The six-point scans give
$d\vp_{\rm H}/d\xi=+2.154\times10^{-3}$ (KS) and $-4.373\times10^{-2}$
(BR) (linear fits; the tangents at $\xi=\tfrac16$ of the quadratic fits
are $+2.156\times10^{-3}$ and $-4.192\times10^{-2}$), while the exact
horizon curvatures are $\mr_{\rm H}r_s^2=+1.574\times10^{-2}$ (KS,
$a/r_s=0.4$) and $-4.062\times10^{-1}$ (BR, $\Omega=0.10$). The slope
ratio, $-20.3$, tracks the parameter-free curvature ratio
$\mr_{\rm H}^{\rm BR}/\mr_{\rm H}^{\rm KS}=-25.8$: two backgrounds whose
horizon curvatures have opposite sign and differ by a factor of $26$ in
magnitude share a common logarithm
$L\equiv(d\vp_{\rm H}/d\xi)/(M^2\mr_{\rm H})$ to $21\%$ ($L=0.547$ versus
$0.431$ at $\mu=\tfrac12$). Table~\ref{tab:xislope} extends the comparison
across the full mass grid, using the $(\xi=0,\tfrac16)$ pairs of the main
survey, and resolves the two regimes of
Eqs.~(\ref{eq:a1finite})--(\ref{eq:slopea2}). At light mass the extracted
$L$ is common to the two geometries to $6$--$7\%$, lies within
$+0.2$--$0.4$ (the measured value of the $O(1)$ constant $c$) of the
zero-parameter logarithm $L_0=2\ln[4\pi T_{\rm H}/(m\,e^{\gamma_E})]$
evaluated at each geometry's own Hawking temperature -- including the
predicted downward shift for the colder BR horizon, whose surface gravity
is $18\%$ below the KS value -- and its logarithmic mass slope is
$dL/d\ln m=-1.87$ for both geometries, against the predicted $-2$. At the
heaviest mass ($\mu=1$, $mr_+\simeq2$) the slope instead matches the
parameter-free large-mass form~(\ref{eq:slopea2}) evaluated with the exact
horizon invariants [$\Box\mr\,r_s^4=-9.54\times10^{-2}$ (KS) and $+1.597$
(BR)]: measured $-1.479\times10^{-2}$ against predicted
$-1.749\times10^{-2}$ for BR ($15\%$), and $+7.74\times10^{-4}$ against
$+9.9\times10^{-4}$ for KS ($22\%$; the smaller and more steeply decaying
curvature signal). At a level comparable to the far-zone $[a_2]$
comparison, the $\xi$-slope is the finite $a_1$-driven
logarithm~(\ref{eq:a1finite}) at light mass, crossing over to the
$\Box\mr$ term of $[a_2]$ at heavy mass, with no fitted parameter beyond
the $O(1)$ constant $c$.

\begin{table}[t]
\centering
\caption{Measured versus predicted $\xi$-slope of the horizon
polarization, expressed as the logarithm
$L=(d\vp_{\rm H}/d\xi)/(M^2\mr_{\rm H})$ of Eq.~(\ref{eq:slopepred}), for
KS ($a/r_s=0.4$, $\mr_{\rm H}r_s^2=+0.01574$) and BR ($\Omega=0.10$,
$\mr_{\rm H}r_s^2=-0.40615$). Provenance of the measured values: the
$m r_s=1$ row is the linear-fit slope of the six-point scan of
Fig.~\ref{fig:xiscan}; every other row is the two-point slope from the
$(\xi=0,\tfrac16)$ pair of the main survey at that mass.
$L_0=2\ln[4\pi T_{\rm H}/(m\,e^{\gamma_E})]$ is the zero-parameter
light-mass logarithm of Eq.~(\ref{eq:a1finite}) (i.e.\ $c=0$), quoted for
$mr_+\lesssim0.5$; $L_{[a_2]}$ is the large-mass form~(\ref{eq:slopea2})
in the same units,
$L_{[a_2]}=[(\bar\xi-\tfrac16)\mr_{\rm H}^2-\tfrac16\Box\mr_{\rm H}]
/(m^2\mr_{\rm H})$ at the scan midpoint $\bar\xi=\tfrac1{12}$, quoted for
$mr_+\gtrsim1$. The $m r_s=0.4$ row sits in the crossover between the two
regimes.}
\label{tab:xislope}
\begin{ruledtabular}
\begin{tabular}{ccccccc}
$m r_s$ & $L^{\rm KS}$ & $L^{\rm BR}$ & $L_0^{\rm KS}$ & $L_0^{\rm BR}$
& $L_{[a_2]}^{\rm KS}$ & $L_{[a_2]}^{\rm BR}$ \\
\hline
0.05 & 5.00 & 4.72 & 4.84 & 4.43 & --- & --- \\
0.1  & 3.70 & 3.42 & 3.45 & 3.05 & --- & --- \\
0.4  & 1.47 & 1.25 & 0.68 & 0.27 & --- & --- \\
1.0  & 0.55 & 0.43 & --- & --- & 1.01 & 0.69 \\
2.0  & 0.20 & 0.15 & --- & --- & 0.25 & 0.17 \\
\end{tabular}
\end{ruledtabular}
\end{table}

\begin{figure}[t]
\centering
\includegraphics[width=\columnwidth]{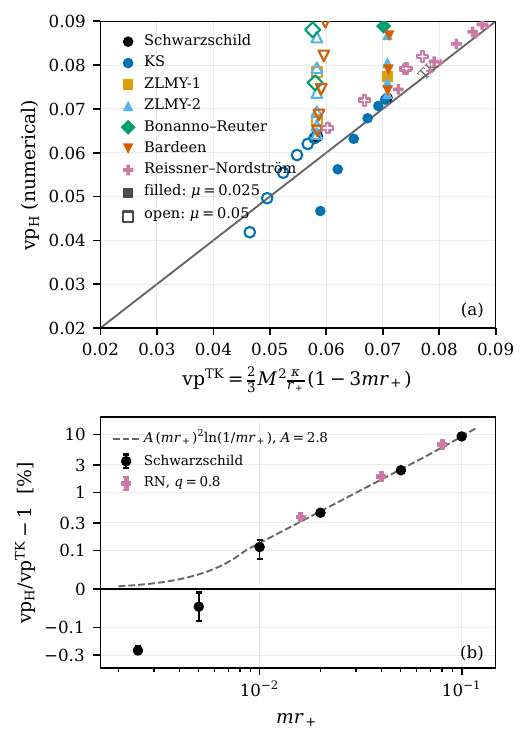}
\caption{(a)~Numerical horizon polarization versus the Tomimatsu--Koyama small-mass
law $\vp^{\rm TK}=\tfrac23M^2(\kappa/r_+)(1-3mr_+)$ evaluated with each
geometry's $(\kappa,r_+)$, for $m r_+\lesssim 0.1$. Filled symbols:
$\mu=0.025$; open symbols: $\mu=0.05$; small filled symbols:
Reissner--Nordstr\"om anchors at $\mu=0.01$; every symbol is a computed
configuration, and the solid diagonal is the TK law itself. The $\mu=0.01$ anchors lie within
$0.5\%$ of the diagonal and at
$\mu=0.025$ ($0.05$) the $\mr=0$ Schwarzschild/RN baseline sits a common
$\approx2\%$ ($7$--$9\%$) above it: in each case this is the truncation offset of the
leading-order law at that mass [panel~(b)]. The deformation contribution beyond
$(\kappa,r_+)$ is the vertical offset from that baseline at matched mass;
offsets exceeding the $\approx2\%$ band around it are significant. (b)~Offset from the TK law, $\vp_{\rm H}/\vp^{\rm TK}-1$, versus
$mr_+$ for Schwarzschild (including additional runs at $mr_+=0.0025$--$0.01$)
and Reissner--Nordstr\"om at $q=0.8$, with the per-point window systematic
as error bars (symmetric-logarithmic vertical axis, linear within $\pm0.1\%$).
Dashed: next-order term $A\,(mr_+)^2\ln(1/mr_+)$, $A=2.8$, fitted to the
Schwarzschild points with $mr_+\ge0.01$. The offset vanishes as $m\to0$.}
\label{fig:tk}
\end{figure}

\begin{figure*}[t]
\centering
\includegraphics[width=\textwidth]{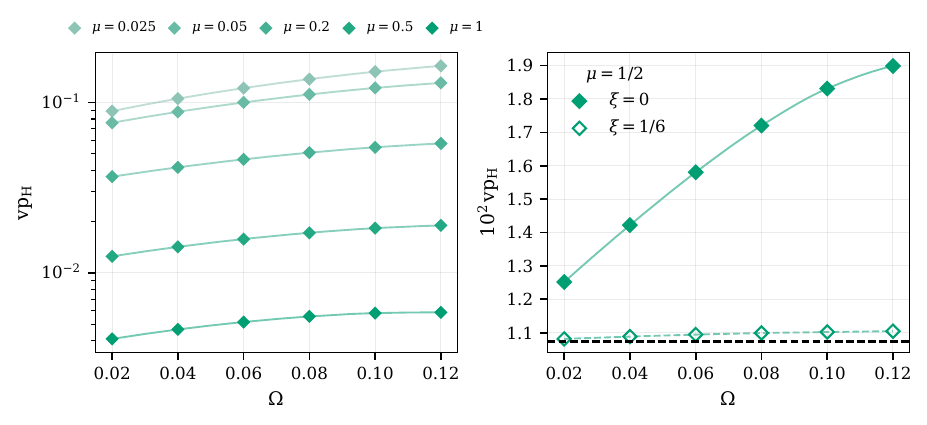}
\caption{Bonanno--Reuter: field-mass dependence ($\mu=0.025$--$1$, left, log
scale, $\xi=0$) and curvature-coupling dependence (right, $\mu=1/2$) of
$\vp_{\rm H}$ versus $\Omega$. Heavier fields are uniformly suppressed --
the light masses show the largest deformation-parameter response -- and conformal
coupling removes the $\Omega$-trend (dashed line: Schwarzschild). Symbols:
computed configurations; thin lines: guide-to-the-eye interpolations.}
\label{fig:massxi_BR}
\end{figure*}

\begin{figure*}[t]
\centering
\includegraphics[width=\textwidth]{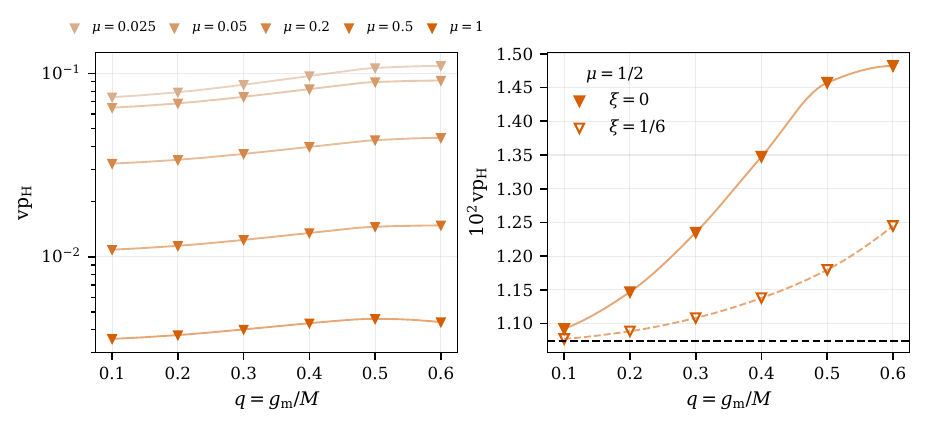}
\caption{Bardeen: field-mass dependence ($\mu=0.025$--$1$, left, log scale,
$\xi=0$) and curvature-coupling dependence (right, $\mu=1/2$) of
$\vp_{\rm H}$ versus the charge $q$. As for BR, heavier fields are uniformly
suppressed and conformal coupling flattens the charge trend (dashed line:
Schwarzschild). Symbols: computed configurations; thin lines: guide-to-the-eye interpolations.}
\label{fig:massxi_Bardeen}
\end{figure*}

\paragraph*{Comparison with the Tomimatsu--Koyama horizon law.}
For Schwarzschild and Reissner--Nordstr\"om, Tomimatsu and
Koyama~\cite{TomimatsuKoyama} derived the analytic small-mass horizon value
$\phisq_{\rm H}=\kappa/(24\pi^2 r_+)\,(1-3mr_+)$, i.e.
$\vp^{\rm TK}=\tfrac23 M^2(\kappa/r_+)(1-3mr_+)$, which depends on the geometry
only through $(\kappa, r_+)$ and reduces at $m=0$ to Frolov's horizon
value~\cite{Frolov:1982} quoted above. Figure~\ref{fig:tk} tests this law against all
seven backgrounds at light masses ($mr_+\lesssim0.1$), using each geometry's own
horizon radius and surface gravity. The $\mr=0$ references (Schwarzschild
and RN) calibrate the comparison: the $\mu=0.01$ RN anchors lie within $0.5\%$ of the diagonal,
while at $\mu=0.025$ and
$0.05$ Schwarzschild and RN sit a common $\approx2\%$ and $7$--$9\%$ above
it. We now show that this offset, including the residual $0.5\%$ at $\mu=0.01$, is the truncation offset of the leading-order (in $mr_+$) law itself, and not a
systematic error of the computation, using additional Schwarzschild runs
down to $mr_+=0.0025$ [Fig.~\ref{fig:tk}(b)]. (i)~The offset vanishes as
$m\to0$: it is $+9.3\%$, $+2.4\%$, $+0.45\%$ and $+0.12\%$ at $mr_+=0.1$,
$0.05$, $0.02$ and $0.01$, and $-0.05\pm0.04\%$ at $0.005$, whereas an
implementation error would generically leave a finite shift. (ii)~For
$0.01\le mr_+\le0.1$ it is described by a single next-order term,
$\vp_{\rm H}/\vp_{\rm H}^{(m=0)}=1-3mr_++A\,(mr_+)^2\ln(1/mr_+)$, with
$A=2.8$ (the value of $A$ extracted point by point varies only between
$2.4$ and $2.8$); Reissner--Nordstr\"om at $q=0.8$ follows the same curve. A positive
$m^2\ln(1/m)$ term of this kind is what one expects at the next order: TK
determine the horizon value only to $O(mr_+)$, noting that the
$O(m^2r_+^2)$ correction requires the mode recurrence to higher
order~\cite{TomimatsuKoyama}; their own low-temperature result,
$8\pi^2r_+^2\phisq_{\rm H}\simeq-m^2r_+^2\,[\tfrac12+\gamma_E+\ln(mr_+)]$
(Sec.~V of Ref.~\cite{TomimatsuKoyama}), is positive for $mr_+<0.34$, and the
flat-space thermal expansion~\cite{DolanJackiw,HaberWeldon} likewise carries
a positive $m^2\ln(T/m)$ term. (iii)~Extrapolating the fit to $m=0$ gives
$\vp_{\rm H}=(0.9990$--$0.9997)\times\tfrac1{12}$, i.e.\ Candelas's exact
massless value to within $0.1\%$. The lightest point, $mr_+=0.0025$, lies
$0.25\%$ below the law, outside its $0.04\%$ window systematic; this
deviation grows rapidly as $m\to0$ and we attribute it, tentatively, to the
loss of precision of the mass-dependent counterterms, which are enhanced as
$1/m^2$ and must cancel against the mode sum. At the masses used elsewhere in
this paper ($\mu\ge0.01$, i.e.\ $mr_+\gtrsim0.013$) its extrapolated size is well below the
other uncertainties of our error budget. The deformation imprint is
therefore the offset from this $\mr=0$ baseline at matched mass, and it
is systematic -- KS falling \emph{below} the baseline
and the negative-$\mr$ geometries \emph{above} it -- so the deviation from the
TK law is a direct, dimensionless measure of the deformation imprint beyond
the classical horizon data $(\kappa,r_+)$, and its sign again tracks the Ricci
scalar.

An additional check concerns TK's \emph{near-extremal} prediction: for massive
fields near an extreme RN horizon they derive a resonance of the horizon
polarization, peaking at $mr_+\simeq0.38$ with asymptotic
($\kappa r_+\to0$) amplitude $Y\equiv8\pi^2r_+^2\phisq_{\rm H}=0.0424$.
Tracking the exact peak along the sequence $q=0.999$--$0.999999$
($\kappa r_+=4.3\times10^{-2}$--$1.4\times10^{-3}$) we find the peak
location converging to the TK value ($mr_+^{\rm peak}=0.31\to0.37$) but the
amplitude climbing only from $Y=0.0292$ to $0.0318$ with visibly decelerating
increments: at every numerically accessible extremality the leading-order TK
amplitude \emph{overestimates} the exact peak by $\gtrsim33\%$ (the exact
peak lies $\gtrsim25\%$ below it), and the
data are consistent with -- though, with only three increments, do not
establish -- saturation below it.\footnote{The sequence advances in
nearly uniform steps of $\log(1/\kappa r_+)$. A slow logarithmic approach to the
TK asymptote would add a roughly constant increment in $Y$ per step; the measured
increments instead shrink ($+0.0011$, $+0.0011$, $+0.0004$), as expected for
convergence to a finite value $Y\simeq0.032$ below the leading-order $0.0424$.
With only three increments available, the last comparable to its own
systematics, we regard this as indicative rather than conclusive; probing one
step deeper is blocked by the $\kappa r_+\sim10^{-3}$ boundary discussed
below.}
Whether the TK limit is
ultimately attained cannot be decided with the present method below
$\kappa r_+\sim10^{-3}$, where the inter-horizon separation falls beneath
the near-horizon window accessible to the mode sum.

\begin{figure*}[t]
\centering
\includegraphics[width=\textwidth]{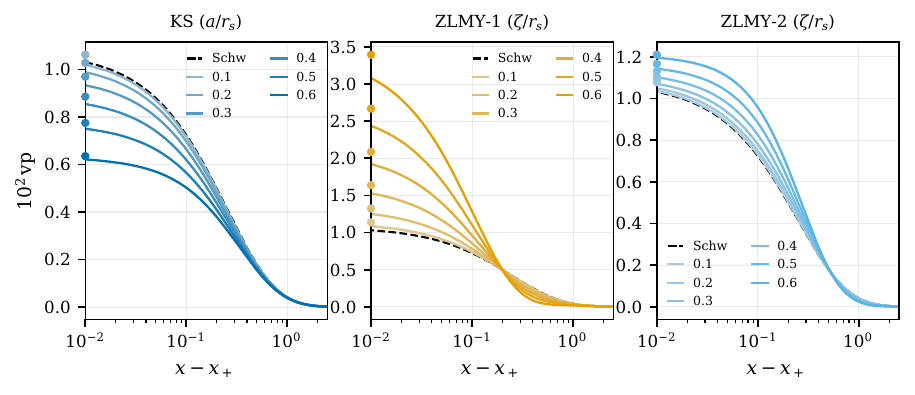}
\caption{Radial profiles $\vp$ versus the distance from the horizon $x-x_+$
(log scale) at $\mu=1/2$, $\xi=0$ for the KS, ZLMY-1 and ZLMY-2 geometries
across the deformation-parameter grid (light to dark with increasing deformation;
Schwarzschild dashed), computed on the full radial grid. The polarization is finite, peaks at the horizon and
decays through the near zone.}
\label{fig:profiles}
\end{figure*}

\paragraph*{Large-$r$ behavior: DeWitt--Schwinger comparison.}
The far-zone profile provides a stringent, parameter-free test of the curvature
interpretation. For a massive field the large-mass DeWitt--Schwinger expansion
gives the scheme-independent leading
term~\cite{Christensen,AHS}
$\phisq_{\rm DS}\simeq[a_2]/(16\pi^2 m^2)$, i.e.
$\vp_{\rm DS}=M^2[a_2]/m^2=\tfrac14[a_2]/m^2$ -- the factor $\tfrac14$ being
$M^2$ in our units $r_s=2M=1$, with $[a_2]$ and $m$ likewise in $r_s=1$
units (for Schwarzschild at $\mu=mM=1$ this yields
$\vp_{\rm DS}=1/(240\,x^6)$, the validation curve below) --
with the heat-kernel coefficient
\begin{align}
[a_2]=&\half\big(\xi-\tfrac16\big)^2\mr^2
      +\tfrac16\big(\tfrac15-\xi\big)\Box\mr \nonumber\\
      &+\tfrac1{180}\big(\mr_{\mu\nu\rho\sigma}\mr^{\mu\nu\rho\sigma}
      -\mr_{\mu\nu}\mr^{\mu\nu}\big),
\label{eq:a2}
\end{align}
evaluated here from the analytic curvature of each background. A qualitative
distinction from the loop-quantum-gravity (Lifshitz-type) case studied in Ref.~\cite{Flachi:2026} emerges. There the background is \emph{not} asymptotically flat,
its Ricci scalar decays only as $\mr\sim r^{-2}$,\footnote{Although $\mr\to0$
pointwise, $r^{-2}$ is precisely the borderline rate: the dimensionless
curvature per sphere $r^2\mr$ tends to a nonzero constant -- an invariant
obstruction no coordinate choice can remove -- and the metric deviation from
Minkowski, roughly the curvature integrated twice, grows as
$\epsilon\ln r$ (indeed $g_{tt}\sim r^{2\epsilon}$, with $\epsilon$ the
quantum-gravity exponent of Ref.~\cite{Flachi:2026}) instead of decaying. The
geometry is locally flat at infinity but not asymptotically flat. For the five
geometries here $\mr$ falls faster than $r^{-2}$, $r^2\mr\to0$, and the metrics
settle onto Minkowski with $1/r$ falloff.} and the slowly-decaying
$\Box\mr<0$ term dominates~(\ref{eq:a2}) at large $r$, producing a genuine
\emph{negative} tail in $\phisq$. For all five geometries considered here
the curvature decays rapidly ($\mr\sim r^{-4}$ for ZLMY-2, $\mr\sim r^{-5}$ for ZLMY-1, BR and Bardeen,
$\mr\sim r^{-6}$ for KS), so that at large $r$ the coefficient~(\ref{eq:a2}) is dominated by the universal,
Schwarzschild-like Kretschmann contribution
$\mr_{\mu\nu\rho\sigma}\mr^{\mu\nu\rho\sigma}\simeq48M^2/r^6>0$. Consequently
$\vp_{\rm DS}$ remains \emph{positive} at large $r$ for every geometry --
including BR, ZLMY-2 and Bardeen, whose Ricci scalar is negative at the horizon (for ZLMY-2 the
coefficient briefly changes sign near $r/r_s\approx2.8$ before the positive
tail sets in) -- and no negative tail develops above the noise floor.
Figure~\ref{fig:dstail} compares the raw numerical $\vp(x)$ with the
parameter-free prediction~(\ref{eq:a2}) at $\mu=1$, $\xi=0$, where the
large-mass expansion is most accurate. The internal angular-truncation
selection of the $\Sigma_2$ mode sum is regularized in $x$, 
so the small tail carries no
discretization artifacts; applied to Schwarzschild the raw output reproduces the
analytic $[a_2]$ curve $\vp_{\rm DS}=1/(240\,x^6)$ to $5\%$ at $r/r_s=3$ and
$0.3\%$ at $3.5$ -- the validation panel of Fig.~\ref{fig:dstail}. The
comparison confirms the DS response: KS and ZLMY-1 agree with the prediction
to $\sim10\%$ over $r/r_s\in[3,4]$ (ratios $1.06/1.08$ and $1.00/1.01$ at
$r/r_s=3.0/3.5$), BR and Bardeen track it with ratios $0.76$--$0.91$; ZLMY-2,
whose deformation signal ($\lesssim5\times10^{-7}$) sits just above the noise
floor, is consistent but noise-limited. Beyond $r/r_s\simeq4$ all curves
approach the intrinsic $\approx3\times10^{-7}$ add-and-subtract floor of the
mode-sum/counterterm cancellation -- present identically in the Schwarzschild
reference -- so the small, geometry-independent negative values at larger $r$
are numerical, not physical.

\begin{figure*}[t]
\centering
\includegraphics[width=\textwidth]{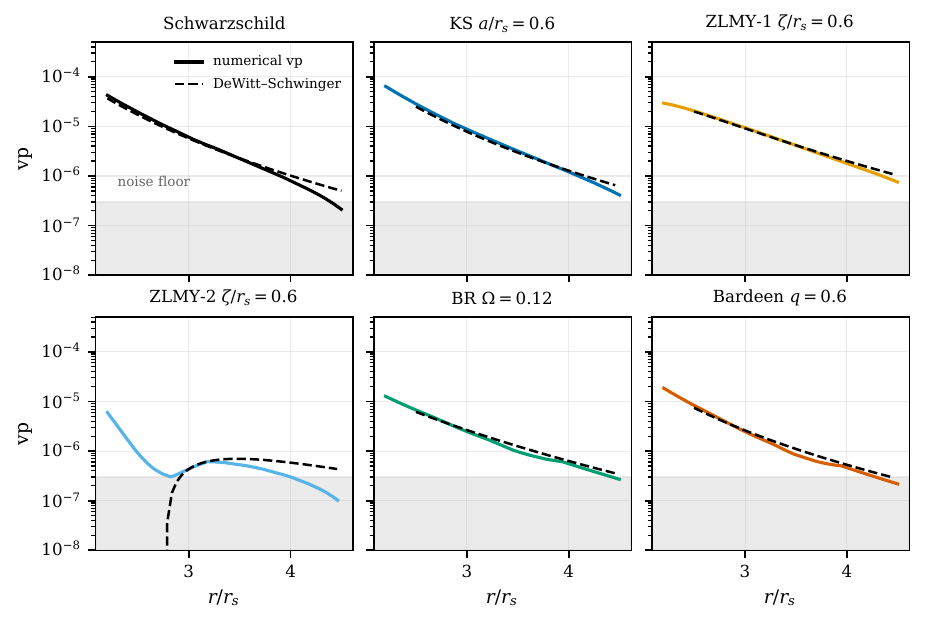}
\caption{Large-$r$ DeWitt--Schwinger comparison at $\mu=1$, $\xi=0$: raw
numerical $\vp$ (solid, no filtering, full radial grid) versus the parameter-free prediction
$\tfrac14[a_2]/m^2$ (dashed), for Schwarzschild (validation panel, analytic
$[a_2]$) and the five geometries at largest deformation. Because all five are
asymptotically flat, $[a_2]$ is Kretschmann-dominated at large $r$ and the
prediction stays positive; the numerical curves track it to $\sim10$--$25\%$
in $r/r_s\in[3,4]$. ZLMY-2 (signal $\lesssim5\times10^{-7}$) is noise-limited.
The shaded band marks the $\sim3\times10^{-7}$ add-and-subtract noise floor.
For ZLMY-2 the DS coefficient changes sign at $r/r_s\approx2.8$; the log axis
suppresses its negative branch, producing the apparent cutoff of the dashed
curve.}
\label{fig:dstail}
\end{figure*}

\section{The near-extremal regime: a sign change of the vacuum polarization}
\label{sec:nearext}
Two of the five geometries -- Bardeen and Bonanno--Reuter -- possess an inner
horizon and an extremal limit ($q\to q_c$, $\Omega\to\Omega_{\rm cr}$) at which the
surface gravity vanishes, giving access to a cold ($\beta=2\pi/\kappa\to\infty$)
Hartle--Hawking regime that the $\zeta$/$a$-type geometries lack. We recall
that the \emph{renormalized} $\phisq$, unlike the manifestly positive
unrenormalized mode sum, is not positive-definite: the subtraction of the
divergent DeWitt--Schwinger terms leaves a finite quantity of either sign,
so a negative value signals no inconsistency. Pushing the
calculation into this regime reveals a qualitatively new phenomenon, which we
did not find for the Reissner--Nordstr\"om black hole in the range probed: at sufficiently large
deformation \emph{and} sufficiently light field mass, the horizon vacuum
polarization becomes \emph{negative}, with $|\vp_{\rm H}|$ exceeding the
massless Schwarzschild horizon value $1/12$ by $(56\pm2)\%$.

\begin{figure}[t]
\centering
\includegraphics[width=\columnwidth]{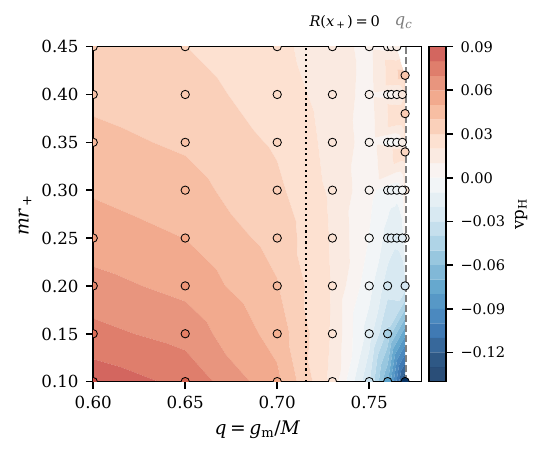}
\caption{Bardeen: sign map of the horizon polarization in the $(q,\,mr_+)$
plane at $\xi=0$. Each circle is one complete mode-sum computation, colored
by its $\vp_{\rm H}$; the background linearly interpolates them. No zero
line is drawn deliberately: the crossing is localized only to within the
near-white zone of the diverging color scale, where
$|\vp_{\rm H}|\lesssim2\times10^{-3}$ is comparable to our uncertainty
and the sign is not resolved (the smallness itself is convergence-checked).
The dotted line marks $q^*=(4/5)^{3/2}$ where the horizon enters the
positive-Ricci core region; the dashed line is the extremal charge $q_c$.}
\label{fig:signmap}
\end{figure}

\paragraph*{The Bardeen sign map.}
Figure~\ref{fig:signmap} maps $\vp_{\rm H}$ over the $(q,\,mr_+)$ plane up to
$q/q_c=0.9996$ ($\beta/r_s=218$). A wedge of negative polarization opens
between $q=0.73$ and $q=0.75$ ($q/q_c\simeq0.95$--$0.97$) and deepens toward
extremality and light mass: at the deep-light corner ($q=0.7695$, $mr_+=0.10$)
$\vp_{\rm H}=-0.130\pm0.002$, i.e.\ $(1.56\pm0.02)\times$ the massless
Schwarzschild horizon value $1/12$ in magnitude (equivalently, $12\times$
the $\mu=1/2$ Schwarzschild reference $1.074\times10^{-2}$). 
For $q\le0.73$ the polarization stays positive down to the lightest masses probed.
The sign boundary $mr_+^c(q)$ is \emph{non-monotonic}: it rises from
$\simeq0.25$ at $q=0.75$ to $\simeq0.35$ in the strip $q\simeq0.76$--$0.765$ -- the maximum extent of
the wedge -- and the crossing falls back to $\simeq0.30$ at $q=0.768$ and to $\simeq0.245$ at $q=0.7695$
(the structure survives a doubling of the mode-sum truncation, and the
optimized and original engines agree to all printed digits). The onset correlates with the horizon Ricci scalar: the Bardeen $R(x)$
changes sign at $x=q$ exactly, the horizon enters the positive-$R$ de~Sitter
core region at $q^*=(4/5)^{3/2}\simeq0.7155$, and the wedge opens once
$R(x_+)r_s^2\gtrsim+0.3$ (between $+0.12$ at $q=0.73$, still positive, and
$+0.44$ at $q=0.75$, negative) -- a KS-like suppression that overshoots into
negative values, with a finite lag past $R(x_+)=0$.

\begin{figure*}[t]
\centering
\includegraphics[width=\textwidth]{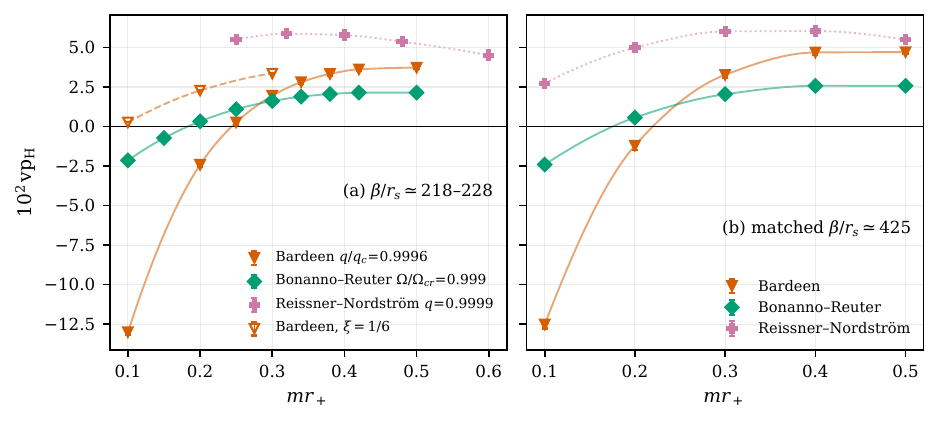}
\caption{Near-extremal mass scans at matched temperature, $\xi=0$.
(a)~$\beta/r_s\simeq218$--$228$: Bardeen ($q/q_c=0.9996$) and Bonanno--Reuter
($\Omega/\Omega_{\rm cr}=0.999$) turn negative at light mass while
Reissner--Nordstr\"om ($q=0.9999$) stays positive with its resonance peak;
open symbols show the Bardeen scan repeated at conformal coupling $\xi=1/6$,
which removes the negative region entirely. (b)~matched
$\beta/r_s\simeq425$: the separation from the singular geometry grows
toward extremality (BR deepens to $-0.024\pm0.001$, the Bardeen value at
$mr_+=0.10$ persists at $-0.125\pm0.003$; RN positive throughout).
Each symbol is an individual computation; the thin lines are
guide-to-the-eye interpolations. Error bars show the per-point window-extrapolation systematic (the
variant set of Sec.~\ref{sec:results}); they are often smaller than the
symbols.}
\label{fig:massscan}
\end{figure*}

What drives the effect? Reissner--Nordstr\"om, which has an inner horizon
and a cold extremal limit but $\mr=0$, provides a control.
Figure~\ref{fig:massscan} shows mass scans at matched temperature. At
$\beta/r_s\simeq218$--$228$ [panel~(a)]: Bardeen and BR both go negative at
light mass (zero crossings $mr_+^c=0.245\pm0.005$ and $0.185\pm0.005$;
deepest sampled values, at $mr_+=0.10$, $-0.130\pm0.002$ and $-0.021\pm0.001$), while RN at the
same depth remains positive everywhere, exhibiting instead the
Tomimatsu--Koyama resonance peak. At matched $\beta/r_s\simeq425$
[panel~(b)] the separation from RN grows: BR deepens to
$-0.024\pm0.001$, the Bardeen value at $mr_+=0.10$ persists at
$-0.125\pm0.003$ (the shift from $-0.130\pm0.002$ is $0.005\pm0.003$, less than twice the window systematic), and RN remains positive throughout.
Matching the inter-horizon separation $x_+-x_-$ instead of the
temperature leads to the same conclusion: RN at $q=0.999$ ($x_+-x_-=0.045$)
and at $q=0.99997$ ($0.0075$), which bracket Bardeen ($0.026$) and BR
($0.025$), gives $\vp_{\rm H}=+0.038$ and $+0.027$ at $mr_+=0.10$. In all our
RN runs, down to $\kappa r_+=4.5\times10^{-3}$, the horizon polarization
stays positive at $\xi=0$, as do the Tomimatsu--Koyama $\kappa r_+\to0$
result~\cite{TomimatsuKoyama} and the massless value~\cite{Frolov:1982}.
In the BR family the wedge opens between
$\Omega/\Omega_{\rm cr}=0.99$ and $0.999$: at $0.99$ the polarization is still
positive but the mass trend has already inverted (rising with $mr_+$), 
and at $0.999$ the crossing sits at $mr_+^c\simeq0.185$; across the same
interval the horizon Ricci scalar changes sign, from $\mr_{\rm H}r_s^2=-0.11$
to $+0.20$. In both regular families, therefore, the negative region appears
only once $\mr_{\rm H}>0$.

Repeating the deepest points at $\xi=1/6$ removes the negative region
entirely [panel~(a), open symbols]: {conformal coupling switches the anomaly off.}
The corner value $-0.130\pm0.002$ becomes $+0.0026\pm0.0001$,
and all twelve near-extremal $(q,\,mr_+)$ combinations probed are positive.
Together with the onset at $\mr_{\rm H}>0$ in both families and the
exact $\xi$-independence of the $\mr=0$ RN control, this identifies the
mechanism as the DeWitt--Schwinger curvature response $\propto(\xi-\tfrac16)\mr$
-- the same term responsible for the moderate-deformation trends of
Sec.~\ref{sec:results}, here driven by a large positive horizon curvature
until it overwhelms the positive thermal contribution. At the Bardeen corner
the slope is $d\vp_{\rm H}/d\xi=+0.80$, i.e.\ $L=+1.7$ per unit
$M^2\mr_{\rm H}$; this is a low-temperature value ($m/T_{\rm H}\simeq39$),
outside the range of Eq.~(\ref{eq:a1finite}), which there would give a
negative $L$. Our data do not establish what produces the positive horizon
curvature. In both families it appears only near extremality, where the
horizon approaches both the de~Sitter core and the inner horizon at once, and
the regular geometries also differ from RN in other respects (a non-vacuum
source in Bardeen, a non-Einsteinian origin in BR); we therefore do not
attribute the effect to the regular core as such.

\section{Conclusions}
\label{sec:conclusions}
We have computed the scalar vacuum polarization $\phisq$ exterior to five
quantum-corrected, effective and regular black holes -- Kazakov--Solodukhin, the two
Zhang--Lewandowski--Ma--Yang loop-quantum-gravity geometries, the
Bonanno--Reuter RG-improved black hole, and the Bardeen regular black hole --
using a metric-agnostic high-performance implementation of the extended point-splitting formalism. To our knowledge these are the first determinations of $\phisq$ for the above mentioned geometries (for the horizonless Bardeen 
the computation has been done recently in Ref.~\cite{BoassoMazzitelli}). The
near-horizon response is geometry-dependent and its sign follows the horizon
curvature: KS, with positive Ricci scalar, suppresses the polarization, whereas
the four negative-curvature geometries enhance it. Measured against the horizon
curvature $M^2\mr_{\rm H}$, which has the same meaning in every theory, the
$\xi$-dependent part of the response is common to all five geometries to within
$\pm20\%$ at light field mass, and at equal curvature the total enhancements of ZLMY-1, BR and
Bardeen agree to within a factor $1.6$; the deformation that leaves the lapse
untouched (ZLMY-2) responds about four times more weakly at equal curvature, and
a $\xi$-independent, geometry-specific part is set by the horizon data
$(\kappa,r_+)$. In every case the result reduces to the
Schwarzschild value in the classical limit, the horizon location, surface gravity
and Ricci scalar are reproduced analytically, and the curvature character of the
imprint -- linear in $(\xi-\tfrac16)\mr$ near the horizon -- is confirmed in
three independent ways: by its strong suppression at conformal coupling, by
the quantitative agreement of the measured horizon slopes
$d\vp_{\rm H}/d\xi$ with the finite $a_1$-driven thermal logarithm at light
mass and with the parameter-free $[a_2]$ form at heavy mass, and by a
parameter-free DeWitt--Schwinger comparison of the far-zone profile. The latter
also draws a sharp line between these geometries and the
loop-quantum-gravity background of Ref.~\cite{Flachi:2026}:
there the slowly-decaying ($r^{-2}$) curvature produces a genuine negative tail in
$\phisq$, whereas here the rapidly-decaying curvature leaves the universal,
positive Schwarzschild contribution in control at large $r$, so no negative
region survives above the noise floor. In the near-extremal regime accessible
to the two geometries with an inner horizon we find a qualitatively new
effect: for sufficiently large deformation and light field mass the horizon
polarization of the \emph{regular} black holes (Bardeen, BR) changes sign,
with $|\vp_{\rm H}|$ exceeding the massless Schwarzschild value $1/12$ by
$(56\pm2)\%$ ($12\times$ the $\mu=1/2$ reference). The sign change appears
only once the horizon Ricci scalar is positive and is removed by conformal
coupling, while Reissner--Nordstr\"om ($\mr=0$) remains positive at matched
temperature and at matched horizon separation -- identifying it as the
DeWitt--Schwinger response to a positive horizon curvature. Whether that
curvature should be attributed to the regular core rather than to the
approach to extremality cannot be decided with these two families. 

Beyond the effective-geometry framework of this paper, the main open
question is how quantum-gravitational modifications of the matter sector,
which enter at the same order in $\hbar$ as the deformation imprint
(Sec.~\ref{sec:geometries}), alter these results; for the loop-quantum-gravity
geometries this requires the corresponding polymer-quantized matter dynamics.
A first natural continuation of this work is to extend the approach used 
for the five solutions considered here plus Schwarzschild and RN, to any other spherically symmetric solution, making the approach truly metric-agnostic. This is in principle possible, but several extensions, particularly regarding the implementation of the boundary conditions are required. A second direction is a closer analysis of the Tomimatsu--Koyama peak, to establish whether the discrepancy found in Sec.~\ref{sec:results} merely reflects the leading-order nature of the TK approximation. Extending this approach to the computation of the renormalized stress tensor
$\langle T_{\mu\nu}\rangle$, which controls back-reaction, is the other open direction; the generality of the
present implementation in the metric functions $f$ and $g$ makes its extension to
further quantum-corrected and regular black holes -- and the assembly of a uniform
$\phisq$ data set across geometries -- feasible. Finally, the results of this work, appropriately extended to a spatially dependent mass, allow us to consider the feedback of the quantum polarization onto the ground state for an interacting quantum field theory on black hole backgrounds. This is a simpler, but physically relevant, version of the quantum back-reaction problem.  

\acknowledgements

This work was conducted using the Miyabi Supercomputing System at the Joint Center for Advanced High Performance Computing (JCAHPC). All data used in this paper is freely accessible on Zenodo~\cite{zenodo_repository}. The larger (11 GB) dataset containing all data generated in the numerical simulation is available upon request. The core Fortran code was written by the author and independently verified against Mathematica and Python implementations; AI tools were used to assist with parameter sweep orchestration, statistical analysis of the data, and generation of the graphics.

\appendix
\section{The numerical engine}
\label{app:code}
This appendix documents the Fortran implementation and records
the design choices that make it fast. The physics content (mode equation,
WKB subtraction, counter-terms) is that of Sec.~\ref{sec:formalism} and of
the closed-form expressions of Refs.~\cite{Flachi:2008sr,Quinta:2016,Flachi:2026};
here we describe how those expressions are organized as a program.
Table~\ref{tab:symbolmap} summarizes, for every ingredient of the
calculation, the evaluation method and the truncation or order
controlling it.

The core engine is written
for the general Euclidean line element
$ds_E^2=x^{2z}f\,d\tau^2+(g/x^2)\,dx^2+x^2\,d\Omega_2^2$, in which an
explicit exponent $z$ multiplies the lapse through the Lifshitz-type
factor $x^{2z}$. The implementation of Ref.~\cite{Flachi:2026} 
considered $z=\epsilon$, i.e. the quantum-gravity
anomalous exponent of its lapse $\sim r^{2\epsilon}$ and fixed the functions $f$ and $g$. Here, we have set $z=0$ and extended the code to general $f$ and $g$ functions of the radial coordinate. $z$ was set to zero at compile time together with the other
parameters, and each $z$-dependent expression below -- the $x^{2z}$ power
tables, $d_A=x^{2z-2}f$, the $(3+z)/x$ term of $c_1$, the indicial
exponent $s=\omega/(x_+^{z}f'(x_+))$ -- reduces to its $z=0$ form. Here, we
keep $z$ explicit because the formulae are transcribed exactly as the
code implements them and can eventually be used for further generalizations beyond those studied in this paper.

\begin{table*}[t]
\centering
\caption{The ingredients of the calculation (symbols of
Secs.~\ref{sec:formalism} and~\ref{app:code}), their evaluation method
and the truncation or order controlling each. Grid arrays have dimension
$N=10^4$; ``spliced'' blocks are generated symbolically per geometry
(Sec.~\ref{sec:implementation}). The $\Delta_k$ are the counterterm
quadratures collected (with the WKB $\zeta$-blocks $P_{1,2,3}$) in $\Xi$
of Eq.~(\ref{eq:phisq}). The functions $\Delta_k$ and $P_{k}$ are the same as those in the Appendix of Ref.~\cite{Flachi:2026}.}
\label{tab:symbolmap}
\footnotesize
\begin{ruledtabular}
\begin{tabular}{llll}
quantity & role & evaluation & truncation / order \\
\hline
radial grid $x_i=x_++e^{u_i}$ & computational mesh in $x=r/r_s$ & uniform in $u=\ln(x-x_+)$ & $N{=}10^4$, $\varepsilon{=}10^{-4}$, $x_{\rm max}{=}200$ \\
$f,\dots,f''''$;\ $g,\dots,g'''$ & metric functions + derivatives & spliced analytic metric block & --- \\
$x_+$;\ $\kappa$, $\beta$, $\alpha$ & horizon; temperature data & scan+bisection+Newton & $4\times10^5$ scan points \\
$\mr$, $\mr'$, $\mr''$ & Ricci scalar + derivatives & closed form in $f,g$ & --- \\
$\sigma,\sigma',\sigma''$;\ $a_{1,2,3}$ & WKB potential; WKB coefficients & closed form, tabulated once & --- \\
$c_2,c_1,c_0$ [Eq.~(\ref{eq:odecoef})] & radial-ODE coefficients & tabulated at $2N{-}1$ RK4 stages & --- \\
$d\ln q_{nl}/dr$,\ $d\ln p_{nl}/dr$ & regular/decaying radial solutions & fixed-step RK4 in $u$ & $N-1$ steps \\
summand of $\Sigma_{1,2}$ [Eq.~(\ref{eq:dif})] & per-mode subtracted term & exact $W_{nl}$ [Eq.~(\ref{eq:Wexact})] $-$ WKB & --- \\
$\Sigma_1$ & static ($n{=}0$) mode sum & blocked $l$-sum, flatness check & $L=2000$ \\
$\Sigma_2$ & thermal double mode sum & checkpointed plateau blend & $n\le150$, $L\le2000$ \\
$\Upsilon_0$ & leading WKB angular sum & quad-precision $l$-sum + tail & $l\le10^4$ \\
$Z_{-1},Z_1,Z_3,Z_5,Z_7$ & resummed WKB $\zeta$-functions & Bessel-$K_\nu$ sums & $n\le200$ \\
$P_1$, $P_2$ & WKB blocks of $\Xi$ & closed form in the $Z$-blocks & --- \\
$P_3=P_{31}{+}P_{32a}{+}P_{32b}$ & WKB block (frequency sum) & Simpson in $\tau$ (qp) + closed forms & $n\le50$, $2000$ nodes \\
$\Delta_{11}$, $\Delta_{12}$ & counterterm frequency sums & Matsubara sums & $n\le2\times10^4$ \\
$\Delta_{21}$, $\Delta_{31}$, $\Delta_{4}$ & counterterm closed forms & closed form & --- \\
$\Delta_{22}$ & counterterm quadrature & midpoint rule on $t\in[0,10]$ & $8000$ nodes \\
$\Delta_{32}$ & curvature counterterm & Simpson; zero at $\xi{=}\frac16$ & $10^4$ nodes \\
$\vp$ & final assembly & $(\pi/\beta)[\Upsilon_0{+}\Sigma_{1,2}{+}P_{1,2,3}{-}\textstyle\sum_k\Delta_k]$ & --- \\
\end{tabular}
\end{ruledtabular}
\end{table*}

\subsection{Layout, precision and build conventions}
The engine is a single self-contained free-form Fortran program of
$\approx2800$ lines with no dependencies beyond the OpenMP runtime. All
procedures are internal (\texttt{contains}) and communicate by host
association: the radial grid and every precomputed geometry table are
program-level static arrays, and \emph{every} numerical control -- grid
size $N=10^4$, truncations, quadrature orders, the physical parameters
$(\mu,\xi,z,r_s)$ and the geometry parameters -- is a compile-time
\texttt{parameter}. This is deliberate: the code-generation front end of
Sec.~\ref{sec:implementation} rewrites these declarations (and the metric
block, below) for each case and recompiles, which takes seconds, so the
compiler sees genuine constants -- array extents are static, loop bounds
fold, and no runtime configuration machinery exists at all. Two real kinds
are used: double precision (\texttt{dp}) throughout, and quadruple
precision (\texttt{qp}, \texttt{selected\_real\_kind(33,4931)}) only in
the few accumulators where a $10^4$-term sum would otherwise lose digits
(Sec.~\ref{app:sums}). The build is
\texttt{gfortran -O2 -fopenmp} with strict IEEE~754 semantics; value-unsafe
optimizations (\texttt{-ffast-math}, architecture-specific fast paths)
are excluded as a hard rule, because the renormalized signal emerges from
cancellations that they demonstrably corrupt. Run environment: unlimited
process stack and \texttt{OMP\_STACKSIZE=64M} (radial work arrays are
thread-private; see Sec.~\ref{app:fast}).

\subsection{Grid, metric block, horizon}
\label{app:grid}
The radial coordinate is compactified as $u=\ln(x-x_+)$: the grid is $N$
uniform points $u_i\in[\ln\varepsilon,\ \ln(x_{\rm max}-x_+)]$ with
$\varepsilon=10^{-4}$, $x_{\rm max}=200$, and $x_i=x_++e^{u_i}$, so the
near-horizon region is resolved exponentially finely.

The geometry enters \emph{only} through nine analytic functions -- $f$
and derivatives to fourth order, $g$ and derivatives to third order --
spliced between marker comments by the symbolic front end. Everything
downstream ($\mr$, $\mr'$, $\mr''$, the WKB potential
$\sigma(x)$ with two derivatives, the WKB coefficients $a_{1,2,3}$, the
ODE coefficients below, the counterterms) is hand-written once, for
general $(f,g,z)$, in terms of these nine functions; no numerical
differentiation of these inputs occurs anywhere (the one exception, a
finite-difference Taylor seed for the static boundary condition, is noted
in Sec.~\ref{app:modes}).

At startup the horizon is located from $f$ alone: a logarithmic scan of
$4\times10^5$ points over $x\in[10^{-3},10^{3}]$ detects the outermost
$(-)\!\to\!(+)$ sign change (dense enough to catch the narrow $f<0$ band
of near-extremal double horizons), followed by bisection and a Newton
refinement to machine precision. The surface gravity is computed generically
as $\kappa=\tfrac12\sqrt{A'(x_+)\,C'(x_+)}$ with $A=g_{tt}$ evaluated by
its analytic derivative. For $C=1/g_{rr}$ the analytic derivative
$C'=2x/g-x^2g'/g^2$ is equally available from the symbolic layer, but in
the engine variable $g=x^2/\tilde g$, which has a pole at the horizon, it
is a removable $\infty/\infty$ at $x_+$ itself; since $C'$ is smooth just
outside, the engine evaluates it at two offsets ($10^{-6}x_+$ and twice
that) and Richardson-extrapolates to $x_+$ -- an $O(h^2)$-accurate step
that reproduces the closed-form surface gravities of
Sec.~\ref{sec:geometries} to $\sim10^{-14}$, the machine-precision
agreement quoted there; $\beta=2\pi/\kappa$.

A single precompute pass then tabulates on the grid every $x$-only
quantity used later: $f,g$ and derivatives, $\mr$, $\sigma,\sigma',
\sigma''$, $a_{1,2,3}$, the ODE coefficients, and auxiliary power tables
($x^{2z}$, $x^{2z+1}$, $x^{2z+2}$, $x^{z+2}$, $\sqrt{g/f}$). All
subsequent hot loops index these tables and re-derive nothing.

\subsection{Radial modes}
\label{app:modes}
For each Matsubara--angular pair $(n,l)$, $\omega=n\alpha$, the radial
equation is written as
$c_2\phi''+c_1\phi'+c_0\phi=0$ with
\begin{align}
c_2&=\frac{x^4}{g},\qquad
c_1=\frac{x^4}{g}\Big(\frac{3+z}{x}+\frac{f'}{2f}-\frac{g'}{2g}\Big),
\nonumber\\
c_0&=-\Big(l(l+1)+\frac{\omega^2}{d_A(x)}+B(x)\Big),
\label{eq:odecoef}
\end{align}
where $d_A=x^{2z-2}f$ and $B=x^2(m^2+\xi\mr)$. The decomposition of
$c_0$ is the key structural point: $c_1$, $c_2$, $d_A$ and $B$ are
independent of both $l$ and $\omega$, so a mode is specified inside the
integrator by just two scalars, $l(l+1)$ and $\omega^2$
(Sec.~\ref{app:fast}). The equation is integrated in the $u$ variable
(state vector $(\phi,\ d\phi/du)$, the transformation contributing the
Jacobian $e^u$ factors) by a \emph{fixed-step} classical fourth-order
Runge--Kutta scheme with $N-1$ steps -- once forward from the horizon
(solution $q$, regular there) and once backward from $x_{\rm max}$
(solution $p$, decaying) -- and only the logarithmic derivatives
$d\ln q/dx$, $d\ln p/dx$ are stored on the grid. Because only
log-derivatives are needed, the integrator renormalizes the state
projectively ($y_2\to y_2/y_1$, $y_1\to\pm1$) whenever
$|y_1|$ leaves $[10^{-100},10^{100}]$, so no mode ever over- or
underflows.

\emph{Boundary conditions.} At $x=x_++\varepsilon$: for $\omega=0$ a
second-order Frobenius expansion of the regular branch,
$\phi=1+a_1^{\rm F}\varepsilon+a_2^{\rm F}\varepsilon^2$ with
$a_1^{\rm F}=-c_0/c_1$ and
$a_2^{\rm F}=-[a_1^{\rm F}(c_1'+c_0)+c_0']/[2(c_2'+c_1)]$, the Taylor
coefficients of $c_{0,1,2}$ obtained by one-sided differences with
offset $\varepsilon/100$ -- the one exception to the all-analytic rule of
Sec.~\ref{app:grid}: these differences act on the already-analytic ODE
coefficients, only seed the $\omega=0$ boundary condition at
$x_++\varepsilon$, and any seed error is damped along the integration
together with the boundary-data transient of
Sec.~\ref{sec:implementation} (they could equally be generated
analytically; the production runs used the form quoted); for $\omega\neq0$ an Eddington--Finkelstein
form $\phi\sim(x-x_+)^{s}(1+b_1(x-x_+))$ with the $l$-independent
indicial exponent $s=\omega/(x_+^{z}f'(x_+))$ ($=n/2$ for Matsubara
frequencies) and the subleading coefficient $b_1$ from the Taylor
expansion of Eq.~(\ref{eq:odecoef}) about $x_+$, giving
$d\ln q(x_++\varepsilon)=s/\varepsilon+b_1$. At $x_{\rm max}$ the
decaying-mode log-derivative of the full equation is imposed in
leading-WKB form, $d\ln p=-P-\sqrt{P^2+V}$ with $P=c_1/(2c_2)$,
$V=-c_0/c_2$, which reproduces the modified-spherical-Bessel
asymptotics at low $l$ and transitions correctly to the centrifugal
regime at high $l$.

A routine \texttt{compute\_dif} assembles from the two integrations the
WKB-subtracted summand of $\Sigma_{1,2}$ at every grid point,
\begin{equation}
d_{l\omega}(x)=\frac{l+\tfrac12}{x^{z+2}}
\left[\frac{(2/x)\sqrt{g/f}}{d\ln q-d\ln p}-\frac{1}{\widetilde W}\right],
\label{eq:dif}
\end{equation}
where the second-order WKB weight is
$1/\widetilde W=\Phi^{-1/2}-\tfrac14\Psi\,\Phi^{-3/2}$ with
$\Phi=[(l+\tfrac12)^2-\tfrac14]f/x^2+\omega^2/x^{2z}+\sigma$ and
$\Psi=a_1\Phi'/\Phi+\tfrac12(a_2-a_3)(\Phi'/\Phi)^2+a_3\Phi''/\Phi$,
evaluated entirely from the precomputed tables. The first bracketed term
is the exact $1/W_{nl}$ of Eq.~(\ref{eq:Wexact}), written in the engine
variable $g=x^2/\tilde g$.

\subsection{Mode sums, plateau truncation, analytic blocks}
\label{app:sums}
$\Sigma_1$ ($n=0$) sums Eq.~(\ref{eq:dif}) over $l=0\ldots L$ with
$L=2000$; the sum is accumulated in blocks of $64$ consecutive $l$,
each block dispatched over OpenMP threads, and a half-$L$ snapshot is
kept so that the flatness of the convergence plateau is reported in the
log.

$\Sigma_2$ ($n\ge1$) is the dominant cost: up to
$n_{\max}=150$ frequencies, each with its own $l$-sum. The $n$-loop is
OpenMP-parallel (dynamic schedule, array reduction); within each $n$ the
$l$-loop terminates early once the mode contributes below $10^{-13}$
everywhere, which makes high frequencies nearly free. The truncation in
$l$ is \emph{not} a fixed cutoff: the cumulative sum converges to a
plateau (physical convergence) and then drifts as the fixed-step RK4
error of the high-$l$ modes accumulates -- the plateau is
grid-independent, the drift is not. The code therefore records the
cumulative sum at $14$ prescribed checkpoints
($L=25,50,80,\ldots,2000$), selects per grid point the flattest adjacent
checkpoint interval, regularizes the resulting integer index field with
a running median (window $201$ points, wider than any observed
misselection block) followed by a boxcar mean (window $121$) into a
continuous fractional index, and evaluates $\Sigma_2$ as the linear
blend of the two bracketing checkpoint columns. This median-blended
selection is what removes the piecewise-constant discretization offsets
mentioned in Sec.~\ref{sec:implementation} and lets the raw tail
reproduce the analytic DeWitt--Schwinger curve with no post-processing.

The static angular block 
$$
\Upsilon_0
=
\sum_l \left[
(l+\tfrac12)/(x^{z+2}\widetilde W)-1/(x^{z+1}\sqrt f)
\right]
$$ 
over $l=0\ldots10^4$. Successive terms cancel to many digits, so this single
sum is accumulated in quadruple precision (OpenMP array reduction over
$l$), and the remainder beyond $l=10^4$ is added analytically: the
large-$l$ expansion of the summand is $O(l^{-2})$ with coefficient
$-x^{1-z}(\sigma_{\rm eff}/2+\Psi_\infty/4)/f^{3/2}$, multiplied by
$\zeta_{\rm tail}=\sum_{l>L}(l+\tfrac12)^{-2}\simeq(L+1)^{-1}
-\tfrac12(L+1)^{-2}$.

The remaining blocks of $\Xi$ are quadratures and closed forms on the
grid, parallelized over grid points: the resummed WKB $\zeta$-functions
$$
Z_q = \sum_n \left(\omega_n^2 + r^{2z} \sigma \right)^{-q/2}
$$
are expressed via the Chowla--Selberg representation \cite{Flachi:2008sr} as sums of modified Bessel functions $K_{0,1,2,3}(2\pi n v)$,
$v=x^z\sqrt\sigma/\alpha$, over $n\le200$ (Numerical-Recipes polynomial
fits; a quadruple-precision Simpson evaluation of the integral
representation is available as a drop-in and confirms the fits at the
$10^{-7}$ level of the noise floor); two closed-form WKB blocks; a
frequency sum of a $\tau$-quadrature (composite Simpson, $2000$ nodes on
$[0,10]$, quadruple-precision accumulator) for the third block, with the
$\tau$-independent factors hoisted out of the quadrature; and the
counterterm pieces -- two Matsubara sums to $n=2\times10^4$, two closed
forms, one midpoint-rule integral ($8000$ nodes; $10^3$ nodes
demonstrably under-resolve the $1/f$-amplified integrand near the
horizon), and one Simpson integral with its removable $t=0$ singularity
replaced by the analytic limit, skipped entirely at $\xi=1/6$ where its
prefactor $(\xi-\tfrac16)$ vanishes identically. The final assembly is
$\vp=(\pi/\beta)\,[\,\Upsilon_0+\Sigma_1+\Sigma_2+P_1+P_2+P_3
-\sum_k\Delta_k\,]$, each block also written to its own plain-text file
for provenance and reuse.

\subsection{What makes it fast}
\label{app:fast}
The optimizations below produced, together, the order-of-magnitude
single-node speedup quoted in Sec.~\ref{sec:implementation}
($\sim\!70$--$90$~s per case at $16$ threads). 

\emph{(a) Precomputed RK4 stage tables.} Since the RK4 steps are fixed
(nodes and midpoints of the $u$-grid) and, by Eq.~(\ref{eq:odecoef}),
the ODE data $(e^u,c_1,c_2,d_A,B)$ are $(l,\omega)$-independent, these
five quantities are tabulated once at the $2N-1$ stage abscissae.
The $\sim\!6\times10^5$ radial integrations of a production run then
perform no metric-function evaluations at all: the innermost RK4 stage
reduces to a handful of arithmetic operations on table entries plus the
two per-mode scalars $l(l+1)$ and $\omega^2$. Two tables are kept,
one per traversal direction, because the forward ($q$) and backward
($p$) sweeps accumulate $u$ with different roundings and bitwise
reproducibility requires honoring each.

\emph{(b) Hoisting of loop invariants.} All $x$-only geometry lives in
grid tables (Sec.~\ref{app:grid}); per-point constants are hoisted out
of Matsubara sums; the $\tau$-independent zero-mode factors are hoisted
out of the $P_3$ quadrature; powers such as $x^{z+2}$ are read from
tables rather than recomputed (a real-exponent power is far costlier
than a multiply).

\emph{(c) Parallelization matched to loop structure.} OpenMP is applied
along two orthogonal axes: over \emph{modes} where each unit of work is
a full radial integration ($\Sigma_1$ in $l$-blocks; $\Sigma_2$ over
$n$, dynamic schedule because early exit makes costs uneven, with array
reductions for the checkpointed sums), and over \emph{grid points}
where the work is pointwise (counterterm sums and quadratures, Bessel
sums, $P_3$; static schedule). Reductions in quadruple precision handle
$\Upsilon_0$. No mutable state is shared: each thread owns its radial
work arrays, which is also why per-call stack footprint was minimized
(the mode-assembly routine uses scalar per-point temporaries instead of
nine grid-sized temporaries, saving $\sim360$~KB per call under
many-thread nesting).

\emph{(d) Doing less work.} The per-$n$ early exit in $\Sigma_2$; the
identically-zero block skipped at conformal coupling; quadrature orders
tuned to the coarsest setting that leaves $\vp$ unchanged at the printed
precision (e.g.\ the $P_3$ $\tau$-quadrature at $2000$ nodes reproduces
the $10^4$-node result exactly while being five times cheaper).

\emph{(e) Compile-time specialization.} Because every size and physical
parameter is a \texttt{parameter} and the metric is spliced source code
(Sec.~\ref{app:grid}), the optimizer constant-folds and vectorizes the
generated expressions per case; regeneration and recompilation cost
seconds and are amortized over minutes of runtime. Notably, plain
\texttt{-O2} with strict IEEE semantics is retained throughout: the
speed comes from the structure above, not from value-unsafe compiler
flags.

Under this organization the run-time budget is dominated by $\Sigma_2$
(the only doubly-summed mode integration), with $\Sigma_1$, $\Upsilon_0$
and all quadratures subdominant.

\end{document}